\documentclass[aps,twocolumn,showpacs,showkeys,amsmath,amssymb,superscriptaddress,nofootinbib,floatfix,longbibliography,eprint]{revtex4-2} 
\usepackage[utf8]{inputenc}
\usepackage{epsfig,bm,slashed,hyperref}

\begin{document}

\title{Gravitational form factors and mechanical properties of the nucleon \\  in a meson dominance approach} 

\author{Wojciech Broniowski}
\email{Wojciech.Broniowski@ifj.edu.pl}
\affiliation{H. Niewodnicza\'nski Institute of Nuclear Physics PAN, 31-342 Cracow, Poland}
\affiliation{Institute of Physics, Jan Kochanowski University, 25-406 Kielce, Poland}

\author{Enrique Ruiz Arriola}
\email{earriola@ugr.es}
\affiliation{Departamento de F\'{\i}sica At\'{o}mica, Molecular y Nuclear and Instituto Carlos I de  F{\'\i}sica Te\'orica y Computacional, 
Universidad de Granada, E-18071 Granada, Spain}

\date{\today}  

\begin{abstract}
We analyze the gravitational form factors and mechanical properties of the nucleon, 
focusing both on some general issues as well as on modeling with meson dominance.
We show that the lattice QCD results for the nucleon gravitational form factors at
$m_\pi=170$~MeV, available for space-like momentum transfer squared up
to $2$~GeV, are explained in a natural way within the meson dominance
approach. We carry out the proper Raman
spin decomposition of the energy-momentum tensor and in each spin channel
use a minimum number of resonances consistent with the perturbative QCD
short-distance constraints. These constraints are related to the
super-convergence sum rules, following from the asymptotic
perturbative QCD fall-off of the form factors. The value of the nucleon $D$-term following from the fits is $-3.0(4)$.
Next, we obtain the two-dimensional transverse gravitational densities of the nucleon in the
transverse coordinate $b$. With the super-convergence sum rules, we
derive new sum rules for these densities at the
origin and for their derivatives, involving logarithmic weighting in the
corresponding spectral density integrals. From analysis of the threshold behavior in
the time-like region and the properties of the $\pi\pi \to N\bar{N}$
reaction, we infer the behavior of the transverse densities at asymptotically large
coordinates. We also carry out the meson dominance analysis of the two- and three-dimensional mechanical
properties of the nucleon (the pressure and stress) and explore their connection to the 
spectral densities via dispersion relations. 
\end{abstract}

\maketitle

\section{Introduction}

The internal structure of the nucleon has been known to exist since the
1950's, when by means of electron scattering it was found that it has a
finite extension, non-trivial magnetic moments, or charge and currents
distributions~\cite{Chambers:1956zz}, which in a relativistically
invariant formulation can be associated with matrix elements of
electromagnetic currents~\cite{drell1962electromagnetic}.  A
remarkable analysis of Frazer and Fulco
\cite{Frazer:1959gy,Frazer:1960zza,Frazer:1960zzb} allowed them to
establish a connection between the electromagnetic nucleon structure and
the $1^{--}$ $\rho$ resonance in the charge form factors of both the pion and
the nucleon, leading to the notion of the vector meson
dominance~\cite{Sakurai:1960ju} and the subsequent current-field
identities~\cite{Lee:1967iv,Kroll:1967it}. In the Breit frame, these
matrix elements are identified as spatial charge and current
distributions~\cite{Sachs:1962zzc,Fleming:1974af}, providing a physically intuitive
picture of a nucleon resembling a droplet made of
charged particles. This image has been ameliorated by
introducing the transverse distributions, which correspond to viewing the
nucleon in the perpendicular direction of a  hadron boosted to
the infinite-momentum
frame~\cite{PhysRevD.15.1141,Burkardt:2000za,Diehl:2002he,Burkardt:2002hr,Miller:2010nz,Freese:2022fat}. 
Then, remarkably, the transverse density of quarks of a particular flavor is manifestly
positive definite~\cite{Burkardt:2002hr,Pobylitsa:2002iu,Diehl:2002he}, allowing for a probabilistic interpretation. 

The considerations of the hadronic matrix elements of the stress-energy-momentum (SEM) tensor 
follow a similar path to the better known EM case, naturally extending to study the energy, momentum, and mass distributions in
terms of the corresponding hadronic form factors. 
Matrix elements of SEM were first
described by Kobzarev and Okun~\cite{Kobzarev:1962wt}. The idea of
tensor meson dominance was introduced and applied by
Freund~\cite{freund1962universality} and by Sharp and
Wagner~\cite{Sharp:1963zz}, when describing the tensor meson exchange
in the NN interaction.  The analytic properties were described by
Pagels~\cite{PhysRev.144.1250}, who suggested saturation by the $2^{++}$
mesons in the dispersion relations (see also~\cite{Schwinger:1965zz}). Current-field identities were
proposed in~\cite{Delbourgo:1966aa,Krolikowski:1967ryy,Raman:1970wq,Raman:1971ur}, 
and a connection to gravity~\cite{patil1967universality,Isham:1971gm} was also given. An early mass radius estimate of the nucleon~\cite{Hare:1972pa}, based on
the sidewise dispersion relations, yielded $0.7~{\rm fm}$.

While many of the relevant issues were essentially understood in the pre-QCD
era, they remained dormant due to the lack of an experimental or lattice
motivation. The situation has changed since Xiang-Dong Ji established the
possibility of relating the deeply virtual Compton scattering (DVCS) to both 
EM current and SEM tensor via the moments of the corresponding Generalized
Parton Distributions (GPDs)~\cite{Ji:1996nm}. 
Later on, Polyakov and Weiss discovered the $D$-term~\cite{Polyakov:1999gs}, and Polyakov
suggested to extend the nucleon picture by invoking its mechanical properties
such as mass, pressure, or shear force distributions,  in an analogy to the classical elasticity 
theory~\cite{Polyakov:2002yz} (see \cite{Ji:2025qax} for an alternative viewpoint), 
analyzed also in~\cite{Lorce:2018egm,Lorce:2021xku,Epelbaum:2022fjc,Panteleeva:2022uii,Lorce:2022cle}.
The mass and spin decomposition in terms of quarks and gluons has been addressed
often, see~\cite{Ji:1995sv,Ji:1996ek,Leader:2013jra,Lorce:2017xzd,Hatta:2018sqd}, with
the proviso that the separation depends on the renormalization scale
and hence is scheme dependent. Asymmetric SEM in the polarized nucleon was considered in~\cite{Won:2025dgc}.

The depiction of the nucleon as a stable droplet in mechanical equilibrium, as well as the relation
of the $D$-term to moments of the pressure and the shear forces, constituting a genuine dynamical 
hadronic property, have triggered a lot of activity over the last years, 
extending the venerable EM current program and methods onto the SEM sector (see, e.g.,~\cite{Burkert:2023wzr} and 
references therein for a review, and~\cite{Lorce:2025oot} for a transparent up-to-date perspective).

Numerous schemes and models have been proposed to describe the gravitational form factors (GFFs). Leading chiral
corrections were addressed in the heavy
baryon~\cite{Belitsky:2002jp,Ando:2006sk,Diehl:2006ya,Moiseeva:2012zi} or 
covariant baryon~\cite{Dorati:2007bk,Alharazin:2020yjv} approaches. Large-$N_c$
scaling was established in~\cite{Goeke:2001tz}.  Estimates were made in the chiral quark
soliton model~\cite{Goeke:2007fp}, the Skyrme
model~\cite{Cebulla:2007ei}, the MIT bag model~\cite{Neubelt:2019sou},
the holographic model~\cite{Abidin:2009hr}, and in AdS/QCD~\cite{Mondal:2015fok,Deng:2025fpq,Fujita:2022jus,Wang:2024sqg}. The QCD sum rules have also been
applied~\cite{Anikin:2019kwi,Azizi:2019ytx}.  Valence quarks in the light front formulation were considered in~\cite{Nair:2024fit},
and a full QCD light front viewpoint was presented in~\cite{Xu:2024sjt}.  Importantly, within perturbative QCD (pQCD), Tong, Ma, and
Yuan~\cite{Tong:2021ctu,Tong:2022zax} have determined the leading asymptotic behavior of GFFs.
A flavor decomposition using light-cone sum rules was presented in~\cite{Dehghan:2025ncw}. Quark and gluon decomposition of the nucleon GFFs was done in~\cite{Ji:2025gsq}, and a soliton calculation with dilaton fields was presented in~\cite{Fujii:2025aip}.

On the experimental and phenomenological side, some constraints have
been obtained via DVCS from CLAS at
JLab~\cite{CLAS:2015uuo,Burkert:2018bqq}, as well as from the
GlueX~\cite{GlueX:2019mkq} data for the $J$/$\psi$
photoproduction~\cite{Wang:2022vhr}.  Along these lines, an estimate
of the proton mass radius~\cite{Kharzeev:2021qkd} was obtained, while
\cite{Du:2020bqj,JointPhysicsAnalysisCenter:2023qgg} questioned
whether the $J$/$\psi$ near-threshold photoproduction could be directly
used to study the nucleon structure. In~\cite{Goharipour:2025lep}, a
determination of GFFs from the Compton form factors was made.
Holographic QCD modeling was carried out in~\cite{Mamo:2022eui}.  A
systematic Bayesian extraction of the proton mass radius based on
photoproduction of vector charmoniums was made in~\cite{Guo:2025jiz}.

The recent  renewed interest in the gravitational
structure of hadrons is spurred on state-of-the-art lattice QCD
calculations involving all the parton
species~\cite{Hackett:2023rif,Hackett:2023nkr,Deka:2013zha,Shanahan:2018pib,Delmar:2024vxn} at low values of
the pion mass, $m_\pi=170$~MeV, close to the physical point.  Similar
studies for the gluon contribution to the trace anomaly form factor
were carried out in \cite{Wang:2024lrm} at higher values of
$m_\pi$. These studies vastly improve on the previous lattice analyses
of the quark GFFs of the
nucleon~\cite{Hagler:2003jd,Gockeler:2003jfa,QCDSF-UKQCD:2007gdl,Hagler:2007hu,LHPC:2007blg,LHPC:2010jcs}. 

In this paper we analyze the GFFs of the nucleon within a purely hadronic
scheme, incorporating the quark-hadron duality principle via
the meson\footnote{\label{f1}In this paper we use the definition of a meson in a broader sense, denoting a baryon-number-zero bound state or resonance, incorporating the quark model 
$q\bar{q}$ states, as well as glueballs or hybrids (see, e.g., Sec.~8.1.1 of~\cite{Gross:2022hyw}).} dominance, supplemented by short-distance constraints from pQCD. These constraints 
imply super-convergence sum rules for the spectral densities associated with GFFs.
The framework provides large-$N_c$ motivated functional expressions in the form of sums over meson pole masses, taken
from the Particle Data Group tables, and the residues fitted to the data.  This
large-$N_c$ scheme, constrained with pQCD, has been successfully used in previous works
for a variety of EM, SEM, and axial form factors, both for the pion and
the nucleon~\cite{Masjuan:2012sk}. Here we extend the method to 
describe the MIT lattice QCD data~\cite{Hackett:2023rif},  using the insights of our
previous analyses for the pion~\cite{Broniowski:2024oyk,RuizArriola:2024udm,Broniowski:2024mpw}.
We use the data for the
full form factors, involving the quark and gluon contributions. That way,
the method is not sensitive to the renormalization scale and scheme, as it
deals with features of conserved currents.

We discuss in detail the two- and three-dimensional spatial
distributions and the mechanical properties of the nucleon coming from
our approach.  We also relate them to the spectral densities via
dispersion relations and obtain novel sum rules for the values of the
above distributions and their derivatives at the origin.  From the
threshold behavior in the time-like region and the analyticity and
unitarity properties of the $\pi\pi \to N\bar{N}$ amplitudes, we find
the forms of the spatial distributions at asymptotically large
coordinates.
  
When this work was in an advanced stage, a paper by Cao, Guo, Li, Yao~\cite{Cao:2024zlf} appeared, presenting
a sophisticated Roy-Steiner equations analysis
incorporating coupled-channel unitarity, analyticity, and crossing for
all channels, including $\pi\pi$, $K \bar K$ and $N \bar N$. The lack
of experimental information for time-like momenta above the $N\bar N$ threshold requires to incorporate contributions from
meson dominance to describe the data of~\cite{Hackett:2023rif}.

\section{Gravitational form factors of the nucleon \label{sec:ff}}

\subsection{Stress-energy-momentum operator \label{sec:general}}

We assume the definition of SEM following from the coupling to gravity, which provides a
symmetric tensor from the outset. In the case of QCD, it coincides with the
Belinfante-Rosenfeld symmetrized SEM~\cite{Collins:1976yq,Pokorski:1987ed} (see also
Appendix~E of~\cite{Belitsky:2005qn}), reading 
\begin{eqnarray}
  \Theta^{\mu\nu} &=& \frac{i}4 \bar \Psi \left[ \gamma^\mu  \overleftrightarrow{D}^\mu +
    \gamma^\nu  \overleftrightarrow{D}^\mu \right] \Psi \nonumber \\
  &-& F^{\mu \lambda a} F^\nu_{\lambda
    a} + \frac14 g^{\mu\nu} F^{\sigma \lambda a} F_{\sigma \lambda a} + \Theta_{\rm GF-EOM}^{\mu\nu},
\end{eqnarray}  
where in the quantized case one has in addition the gauge-fixing and
the equations-of-motion terms. Under a Lorentz transformation $ x^\mu \to
\Lambda^\mu_\alpha x^\alpha$, it transforms covariantly,
$\Theta^{\mu\nu} (x) \to \Lambda^\mu_\alpha \Lambda^\nu_\beta
\Theta^{\alpha\beta} (\Lambda^{-1} x)$, but not irreducibly. A naive
decomposition into a traceless and traceful pieces
\begin{eqnarray}
\Theta^{\mu\nu} = \Theta_{S,n}^{\mu \nu} + \Theta_{T,n}^{\mu \nu}   \equiv \frac14 g^{\mu\nu} \Theta + 
\left[ \Theta^{\mu\nu} - \frac14 g^{\mu\nu} \Theta \right] ,
\end{eqnarray}
with $ \Theta = \Theta ^\mu_\mu$, {\it does not} comply to a separate conservation of the 
two components, since $\partial_\mu \Theta_{S,n}^{\mu \nu} =-\partial_\mu \Theta_{T,n}^{\mu \nu}  \neq 0$.
In particular, this becomes a problem in the application of the meson dominance.  
A consistent 
decomposition where the two tensor components are conserved separately was proposed long ago by Raman~\cite{Raman:1971jg}, 
$\Theta^{\mu \nu} = \Theta_S^{\mu \nu} + \Theta_T^{\mu \nu}$, with
\begin{eqnarray}
\Theta_S^{\mu \nu} = \frac13 \left[g^{\mu \nu} - \frac{\partial^\mu \partial^\nu }{\partial^2} \right] \Theta 
\implies  \partial_\mu \Theta_S^{\mu \nu} =\partial_\mu \Theta_T^{\mu \nu} =0. \nonumber \\ 
\end{eqnarray}
It was recently also used in~\cite{Fujita:2022jus,Sugimoto:2025btn}.
The two irreducible tensors have a 
well-defined angular momentum,  $J^{PC}=0^{++}$  and $2^{++}$, correspondingly.
Although this decomposition involves an apparent nonlocality, it does not lead to issues in our approach.

\subsection{Matrix elements}

With the symmetric SEM operator, 
the three gravitational form factors of the nucleon are defined through the matrix element in on-shell nucleons, 
\begin{eqnarray}
&&    \langle p^\prime,s^\prime| \Theta_{\mu\nu}(0) |p,s\rangle
    = \bar u(p',s') \biggl[
      A(t)\,\gamma_{\{\mu} P_{\nu\}}  \nonumber \\
&&    + B(t)\,\frac{i\,P_{\{\mu}\sigma_{\nu\}\rho}q^\rho}{2m_N}
    + D(t)\,\frac{q_\mu q_\nu-g_{\mu\nu}q^2}{4m_N} \biggr]u(p,s),
    \label{eq:defwB}
\end{eqnarray}
where $m_N$ is the nucleon mass, the Dirac  
spinors are normalized as $\bar u(p,s)\, u(p,s) =2m_N$, 
$P=(p'+p)/2$, $q=p'-p$,  $t=q^2$, $\sigma_{\mu\nu}=\tfrac{i}{2}[\gamma_\mu,\gamma_\nu]$, and  $a_{\{\mu} b_{\nu\}}=\tfrac{1}{2}(a_\mu b_\nu + a_\nu b_\mu)$. Clearly, $P\cdot q=0$ and $P^2=m_N^2-t/4$. The $A(t)$ form factor is chirally even, whereas
$B(t)$ and $D(t)$ are chirally odd.

Through the use of the Gordon identity
\begin{eqnarray}
2m\bar u^\prime\gamma^\alpha u=\bar u^\prime(2P^\alpha+i\sigma^{\alpha\rho}q_\rho)u, \label{eq:gordon}
\end{eqnarray}
the definition~(\ref{eq:defwB})
can be equivalently written in different ways, which due to the mass term
may mix the chiral properties. For instance, 
\begin{eqnarray}
&&    \langle p^\prime,s^\prime| \Theta_{\mu\nu}(0) |p,s\rangle
    = \frac{1}{m_N} \bar u' \biggl[
      A(t) P_\mu P_\nu\  \nonumber \\
&&    + J(t)\,i\,P_{\{\mu}\sigma_{\nu\}\rho}q^\rho
    + D(t)\,\frac{q_\mu q_\nu-g_{\mu\nu}q^2}{4} \biggr]u.
    \label{eq:defwJ}
\end{eqnarray}
The form factors $A$, $B$, and $J$ are related,
\begin{eqnarray}
J(t)=\frac{A(t)+B(t)}{2}. \label{eq:J}
\end{eqnarray}
Another form of SEM that can be written is 
\begin{eqnarray}
&&    \langle p^\prime,s^\prime| \Theta_{\mu\nu}(0) |p,s\rangle
    = \bar u' \biggl[
      2J(t)\,\gamma_{\{\mu} P_{\nu\}}  \nonumber \\
&&    - B(t)\frac{P_\mu P_\nu}{m_N}
    + D(t)\,\frac{q_\mu q_\nu-g_{\mu\nu}q^2}{4m_N} \biggr] u.
    \label{eq:defwBalt}
\end{eqnarray}
Normalizations are $A(0)=1$ and $J(0)=\tfrac{1}{2}$ (this last
condition is referred to as Ji's sum rule~\cite{Ji:1996nm}). These
imply $B(0)=0$, which is known as the vanishing of the ``anomalous
gravitomagnetic moment'' of the nucleon
\cite{Kobzarev:1962wt}. Finally, the value $D(0)$ is the $D$-term
\cite{Polyakov:1999gs}, which is a dynamical quantity to be extracted
from {\it ab initio} calculations or data (similarly to the magnetic
moments or the axial coupling constant).

The trace of SEM, corresponding to the divergence of the dilation current, $J_\mu^D = \Theta_{\mu \nu}x^\nu $,
and the trace anomaly of QCD, $\Theta_\mu^\mu=\partial^\mu J^D_\mu$, 
are related to the scalar gravitational form factor $\Theta(t)$, 
\begin{eqnarray}
 &&\langle p^\prime,s^\prime| T_\mu^\mu(0) | p,s\rangle = \bar u(p',s') \Theta(t) \bar u(p,s), \nonumber \\
 &&\Theta(t) = \frac{1}{m_N}\left[ (m_N^2 -\frac{t}{4}) A(t) - \frac{3}{4} t D(t) + \frac{1}{2} t J(t) \right ],  \label{eq:Th}
\end{eqnarray}
with the condition $\Theta(0)=m_N$. From there one has
\begin{eqnarray}
  D(0)= \frac{4 m_N}{3} \left[ m_N A'(0)-\Theta'(0) \right] \, . 
\end{eqnarray}
The extension of the Raman decomposition introduced in Sec.~\ref{sec:general} to the nucleon case is obtained by 
explicitly choosing a mutually orthogonal basis in the tensor components. For the matrix element 
$ \Theta^{\mu \nu} \equiv \langle p^\prime,s^\prime| T^{\mu \nu}(0) | p,s\rangle$ it takes the form 
\begin{eqnarray}
&& \Theta^{\mu \nu} =  \Theta_S^{\mu \nu}+\Theta_T^{\mu \nu}, \\
&& \Theta_S^{\mu \nu} = \frac13 Q^{\mu \nu} \bar u^\prime\Theta(t) u, \nonumber \\ 
&&  \Theta_T^{\mu \nu} = \Theta^{\mu \nu}- \frac13  Q^{\mu \nu} \bar u^\prime\Theta(t)  u,  \nonumber 
\end{eqnarray}
where the transverse tensor is defined as 
\begin{eqnarray}
Q^{\mu \nu}\equiv g^{\mu \nu}-{q^\mu q^\nu}/{q^2},   \label{eq:Q}
\end{eqnarray}
fulfilling $Q^\mu_\mu= 3$ and $q^\mu Q_{\mu\nu}=0$. 
Explicitly,
\begin{eqnarray}
&&  \Theta_T^{\mu \nu}= \frac{\bar u^\prime u}{m_N} \left[ P^\mu P^\nu - \frac{P^2}3  Q^{\mu \nu} \right] A(t), \label{eq:ram2}\\
&& + \frac{1}{m_N}\bar u^\prime\left [ \,i\,P^{\{\mu}\sigma^{\nu\}\rho}q_\rho -\frac{t}{6} Q^{\mu\nu} \right ] u \,J(t), \nonumber
\end{eqnarray} 
Thus, as already stressed, the Raman decomposition implements a separate conservation of
the scalar and tensor parts, namely $ q_\mu \Theta_S^{\mu\nu}=q_\mu
\Theta_T^{\mu\nu}= 0$. We note that the apparent non-locality in Eq.~(\ref{eq:Q}) does not plague 
the considered quantities, as we only use the properties $q_\mu Q^{\mu \nu}=0$ and $Q^\mu_\mu=3$. 
 The separate pole terms in $1/q^2$ cancel in
the total sum by the condition $\Theta(0)= m_N A(0)$. Moreover, the
$A$ and $J$ terms of Eq.~(\ref{eq:ram2}) are separately traceless and
conserved.  Equation~(\ref{eq:ram2}) can be cast in a form involving
mutually orthogonal, traceless, and conserved tensors, namely
\begin{eqnarray}
&&  \Theta_T^{\mu \nu}= \frac{\bar u^\prime u}{m_N} \left[ P^\mu P^\nu - \frac{P^2}{3}  Q^{\mu \nu} \right] \left( A(t)+\frac{t}{2P^2} J(t)\right) \nonumber \\
&& + \frac{1}{m_N}\bar u^\prime\left [ \,i\,P^{\{\mu}\sigma^{\nu\}\rho}q_\rho -\frac{t}{2P^2}  P^{\mu}P^{\nu} \right ] u \,J(t), \label{eq:ram3} 
\end{eqnarray} 
which has interesting unitarity properties in the time-like region, corresponding to the helicity non-flip and flip amplitudes
(see Sec.~\ref{sec:spec-low}).   

From what has been said, $\Theta$ is the form factor of the scalar $0^{++}$ (trace anomaly) channel, 
while $A$ and $J$ correspond to the tensor channel $2^{++}$. On the other hand, $D$ mixes the spin quantum numbers and in that sense is 
less fundamental. From Eq.~(\ref{eq:Th})
\begin{eqnarray}
\hspace{-3mm} D(t)=\frac13 \left [ \frac{4 P^2}{t} A(t)+2 J(t) -\frac{4m_N}{t} \Theta(t)  \right ]. \label{eq:Drel}
\end{eqnarray}
The decomposition described above is needed for the meson saturation model, as it determines which mesons should be used for a particular form factor. We will therefore use $\Theta$, $A$, and $J$ as the basic form factors and obtain $D$ from Eq.~(\ref{eq:Drel}).
Alternatively, one can use
\begin{eqnarray}
\hspace{-3mm} - t D(t)=\frac13 \left [ 4 m_N^2 \left (\frac{\Theta(t)}{m_N} - A(t) \right ) - t B(t)  \right ], \label{eq:Drel2}
\end{eqnarray}
which is the quantity appearing in the analysis of the mechanical properties.

With identity~(\ref{eq:Drel}), the decomposition~(\ref{eq:defwBalt}) can also be written in the form involving terms which are of good spin and separately 
conserved, which will become useful in Sec.~\ref{sec:unit}, namely
\begin{eqnarray}
&&    \langle p^\prime,s^\prime| \Theta_{\mu\nu}(0) |p,s\rangle
    = \bar u' \biggl[
      2J(t) \left ( \gamma_{\{\mu} P_{\nu\}} - \frac{1}{3} m_N Q_{\mu\nu} \right )\nonumber \\
&&    - B(t)\frac{1}{m_N} \left ( P_\mu P_\nu -\frac{P^2}{3} Q_{\mu \nu} \right )
    + \Theta(t) \frac{1}{3} Q_{\mu\nu} \biggr] u.
    \label{eq:defwBalt2}
\end{eqnarray}
The tensors multiplying the spin-2 form factors are conserved and traceless. 

\subsection{Asymptotics from pQCD \label{sec:pQCD}}

Within pQCD, Tong, Ma, and Yuan~\cite{Tong:2021ctu,Tong:2022zax} derived the asymptotic formulas for the GFFs on the nucleon 
at large $-t$, which are
\begin{eqnarray}
&& A(t) \sim + \frac{\alpha(t)^2}{(-t)^2}, \;\;\;\; J(t) \sim + \frac{\alpha(t)^2}{(-t)^2} \nonumber \\
&& B(t) \sim - \frac{\alpha(t)^2}{(-t)^3},  \;\;\;\; D(t) \sim - \frac{\alpha(t)^2}{(-t)^3}. \label{eq:asy}
\end{eqnarray}
Here $\alpha(t)$ is the running strong coupling constant, which at the leading
order reads
\begin{eqnarray}
\alpha(t)=\frac{4\pi}{\beta_0 \ln(-t/\Lambda_{\rm QCD}^2)},
\end{eqnarray} 
with $\beta_0=(11N_c-2N_f)/3=9$ for three active flavors.  In
Eq.~(\ref{eq:asy}), $\pm$ indicates the overall sign (we do not attempt
to give the values of the constants, which depend on the specific
nucleon partonic distribution amplitudes (DAs)).  As expected, the chirally
odd form factors $B(t)$ and $D(t)$ fall off with an extra power of $t$
as compared to the chirally even ones, $A(t)$ and $J(t)$. Actually, it
has been found that the non-trivial relation $B(t)=3 D(t)$ holds asymptotically to
leading order in pQCD~\cite{Tong:2021ctu,Tong:2022zax}.

For the scalar (trace anomaly) form factor $\Theta(t)$, the situation is somewhat more complicated. From
Eq.~(\ref{eq:Th}) $ m_N \Theta(t) = m_N^2 A(t) + \frac{t}{4} (
B(t)-3 D(t)) ={\cal O}( m_N^2 \alpha(t)^2 /t^2)$. However, the pQCD hard kernel
behaves as $\alpha(t)^3/(-t)^2$, and the presence of the soft
end-point singularities in the nucleon DAs alters this behavior. In particular, Eq.~(5.15) in~\cite{Tong:2022zax}, applying 
the DAs from~\cite{Braun:2006hz}, 
reads
\begin{eqnarray}
\Theta(t) \sim - \frac{\alpha(t)^3}{(-t)^{3/2}}\log^2 \left( \frac{-t}{\Lambda_c^2} \right), \label{eq:thasy}
\end{eqnarray}
where $\Lambda_c\simeq 200$~MeV is a scale 
associated with DAs. Note a negative sign asymptotically. As concluded in~\cite{Tong:2022zax}, since 
$\Theta(0) = m_N>0$,  Eq.~(\ref{eq:thasy}) requires a change of sign in $\Theta(-t)$ in the space-like domain $-t>0$. 
Equations (\ref{eq:asy}) are consistent, up to a weak logarithmic
dependence, with the generic QCD counting
rules~\cite{Alvegard:1979ui,Brodsky:1980sx}. 

\subsection{Dispersion relations and sum rules}

Quite generally, the nucleon form factors $F(t)$ satisfy analytic
properties: They are real in the space-like region, $t=-Q^2 \le 0$,
and possess a branch cut discontinuity starting at  $s \ge 4 m_\pi^2$,
corresponding to the $ \pi\pi \to N \bar N $ process below the $N \bar
N$ threshold $s= 4 m_N^2$. Hence, up to possible subtractions, a form
factor $F(-Q^2)$ satisfies the dispersion relation
\begin{eqnarray}
F(-Q^2)=\frac{1}{\pi} \int_{4m_\pi^2}^\infty ds \frac{{\rm Im}F(s)}{s+Q^2}, \label{eq:dis}
\end{eqnarray}
where $Q^2=-t$ is the space-like momentum transfer squared. 

The integrals in Eq.~(\ref{eq:dis}) converge if asymptotically ${\rm Im}\, F(s)$ tends to zero sufficiently fast, which is the case in QCD. 
With the conditions~(\ref{eq:asy}), one 
finds immediately (by expanding $1/(s+Q^2)=1/Q^2-s/Q^4+\dots$) the so-called super-convergence sum rules~\cite{Mergell:1995bf,Donoghue:1996bt,Belushkin:2006qa,Broniowski:2024oyk},
\begin{eqnarray}
&& \lim_{Q^2 \to \infty}Q^{2n}F(Q^2) = 0 \;\;\;\; \Rightarrow \;\;\;\; \int_{4m_\pi^2}^\infty ds \,s^{n-1}{\rm Im}F(s) =0, \nonumber \\
&& \hspace{5.5cm} n=1, 2, \dots . \label{eq:srn}
\end{eqnarray}
This set of sum rules must be satisfied by the spectral density ${\rm Im}F(s)/\pi$ of a given form factor. In particular, from~(\ref{eq:asy}) it follows that 
from  pQCD $A$ and $J$ satisfy the sum rules with $n=1, 2$, while $B$ and $D$ with $n=1, 2, 3$. The presence of $\alpha$, carrying additional damping, is needed for the 
highest $n$ value. Indeed, the time-like large-$s$ behavior corresponding to~(\ref{eq:asy}) is 
\begin{eqnarray}
&&  {\rm Im}\,A(s) \sim + \frac{1}{s^2 L^3}, \;\;\;\; {\rm Im}\,J(s) \sim +  \frac{1}{s^2 L^3}, \nonumber \\
&&  {\rm Im}\,B(s) \sim + \frac{1}{s^3 L^3},  \;\;\;\; {\rm Im}\, D(s) \sim + \frac{1}{s^3 L^3}, \label{eq:asys}
\end{eqnarray}
with the meaning ${\rm Im}\,A(s)\equiv {\rm Im}\,A(s+i \epsilon)$. In the derivation
above, we have used the analytic continuation into the complex $t=s$ plane, $-t=\exp(-i \theta)|t|$, whence
$\alpha(s + i \epsilon)={4\pi}/({\beta_0 (L-i \pi)})$, with the short-hand notation $L=\ln(s/\Lambda^2)$.

For the case of $\Theta$, 
\begin{eqnarray}
{\rm Im}\,\Theta(s) \sim - \frac{1}{s^{3/2}} \frac{L_c^2}{L^3}, \label{eq:asysTh}
\end{eqnarray}
where $L_c=\log ( s/ \Lambda_c^2 )$.
Therefore ${\rm Im}\,\Theta(s)$ satisfies the sum rule~(\ref{eq:srn}) with $n=1$ only. 

Since all the considered GFFs of the nucleon satisfy~(\ref{eq:srn}) with $n=1$, namely, $\int_{4m_\pi^2}^\infty ds \, {\rm Im}F(s) =0$, all 
the functions ${\rm Im}F(s)$ cannot have a definite sign in the time-like domain 
$4m_\pi^2 \le s <\infty$. This feature is analogous to the case of the gravitational spectral densities of the pion~\cite{Broniowski:2024oyk}.

The charge-type sum rules have the form 
\begin{eqnarray}
F(0)=\frac{1}{\pi} \int_{4m_\pi^2}^\infty ds \frac{{\rm Im}F(s)}{s}, \label{eq:charge}
\end{eqnarray}
whereas the derivatives with respect to $t$ at the origin are given by
\begin{eqnarray}
\left . \frac{d^nF(t)}{dt^n} \right |_{t=0}=\frac{n!}{\pi} \int_{4m_\pi^2}^\infty ds \frac{{\rm Im}F(s)}{s^n}, \label{eq:slope}
\end{eqnarray}
from where, as is well known, it follows that precise modeling at low $t$ requires a detailed knowledge of the spectral densities near the $2\pi$ production threshold.

\subsection{Spectral densities at threshold
\label{sec:spec-low}}

For our later analysis it is pertinent to determine some qualitative
features of the spectral densities. In what follows, we will show that
in the region $ 4m_\pi^2 \le t \le 4m_K^2 $ (between the $2\pi$ and $K
\bar K$ thresholds) the signs of the spectral functions corresponding
to the Raman form factors $\Theta$, $J$, $A + t J /2P^2 = m_N^2/P^2[ A +
tB/4 m_N^2]$, and  $B$ are well defined and in
fact positive. This is consistent with positivity in the space-like region
at small $t=-Q^2$. 

As already mentioned, the spectral functions 
corresponding to the nucleon GFFs start at $t=4 m_\pi^2$.
They reflect the leading singularity in the $\pi \pi \to N \bar
N$ process, which is not observable {\it below} the $N \bar N $
threshold $t= 4m_N^2$, but can be related to the observable pion form
factors via unitarity and analyticity.  The unitarity relations for
the GFF were derived in~\cite{Cao:2024zlf} and can be
written in several useful equivalent forms, for instance
\begin{eqnarray}
 && {\rm Im} \,\Theta (t) =  \frac{3 \sigma_\pi f_{0,+} (t) \Theta_\pi^* (t)}{2(4m_N^2-t)},  \\  
 && {\rm Im } \,J (t) = \frac{3 t^2 \sigma_\pi^5}{64 \sqrt{6}} f_{2,-} (t) A_\pi^*(t), \nonumber  \\
  &&  {\rm Im }\,A (t) + \frac{2 t {\rm Im} J(t)}{4 m_N^2 -t} =  \frac{3 t^2 \sigma_\pi^5 m_N}{16 (4m_N^2-t)}  f_{2,+} (t) A_\pi^*(t), \nonumber
\end{eqnarray}
where $\sigma_\pi= \sqrt{1-4 m_\pi^2/t}$. Note that the combinations corresponding to the mutually orthogonal Raman-like decomposition of Eq.~(\ref{eq:ram3}), involving the form factors  
$J(t)$ and $A(t) + t J(t) /2P^2 = m_N^2/P^2 [A(t) +  tB(t)/4 m_N^2 ]$, comply naturally with the spin flip and spin non-flip
$N \bar N$ amplitudes. From the above relations we immediately get
\begin{eqnarray}
  {\rm Im}\, B(t) &=&  \frac{3 t^2 m_N \sigma_\pi^5}{16 (4m_N^2-t)} \nonumber \\
               &\times & \left[ \frac{m_N \sqrt{2} f_{2,-}(t)}{\sqrt{3}}  - f_{2,+}(t) \right] A_\pi^*(t). 
\end{eqnarray}

From Watson's theorem for the pion GFFs one has 
\begin{eqnarray}
\Theta_\pi(t) &=& |\Theta_\pi(t)| e^{i \delta_{00}(t)}, \nonumber \\
A_\pi(t) &=& |A_\pi(t)| e^{i  \delta_{02}(t)},
\end{eqnarray}  
 where $\delta_{00}$ and $\delta_{02}$ are the isoscalar S- and D-wave
 elastic scattering phase shifts for the $\pi\pi \to \pi \pi $ process,
 respectively. While the elastic scattering takes place strictly below the
 $\pi\pi$ production threshold, i.e., for the process $2 \pi \to 4 \pi$
 starting  at $s=16 m_\pi^2$, it has been found over the years that in
 practice the inelasticity takes off at the $K \bar K$ threshold $s=4m_K^2$.

Clearly, since ${\rm Im}\, \Theta (t)$ is a real quantity, $f_{0,+} (t) = \pm |f_{0,+} (t)| e^{i \delta_{00}(t)}$, where the sign
can be fixed by analyzing the $\pi N \to \pi N $ data in the physical
region $s \ge (m_\pi+ m_N)^2$ and analytically continuing to the
unphysical region using unitarity and crossing. The net result is a
rapid change of $|f_{0,+} (s)|$ around and above the threshold region,
due to a subthreshold singularity at $s= 4m_\pi^2 -m_\pi^4/m_N^2$
stemming from the nucleon pole exchange contribution. 

In the case of
the D-wave form factors $A$ and $J$, the result involves both the spin flip
and non-flip $N \bar N$ scattering amplitudes. Similarly to the S-wave case, up to a
sign they have a phase identical to the isoscalar D-wave
$\pi\pi$ scattering phase shift, $\delta_{20}(t)$, such that
$f_{2,\pm}(s) = \pm |f_{2,\pm}(s)| e^{i \delta_{20}(s)}$. The values
around the threshold have been determined in the venerable
Karlsruhe-Helsinki (KH80) analysis~\cite{Koch:1980ay} (numerical
values are tabulated in \cite{hohler1983pion}) and reviewed with
uncertainties by the Bonn group~\cite{Hoferichter:2015hva}.
The data analysis supports the plus sign in all channels, such that 
in the interval
$4m_\pi^2 \le t \le 4 M_K^2 $ one has
\begin{eqnarray}
&&  {\rm Im} \,\Theta (t) =  \frac{3 \sigma_\pi |f_{0,+} (t)|| \Theta_\pi (t) |}{2(4m_N^2-t)}  >0,  \\  
&&  {\rm Im } \,J (t) = \frac{3 t^2 \sigma_\pi^5}{64 \sqrt{6}} |f_{2,-} (t) || A_\pi(t) | >0, \nonumber \\
&&    {\rm Im }\,A (t) + \frac{2 t {\rm Im} J(t)}{4 m_N^2 -t} =  
  \frac{3 t^2 m_N\sigma_\pi^5}{32 \sqrt{6}} | f_{2,+} (t) || A_\pi(t) |>0, \nonumber  \\
&&  {\rm Im} \,A(t) =   \frac{3 t^2 \sigma_\pi^5 |A_\pi(t) |}{16 (4m_N^2-t)} \! \left[ m_N |f_{2,+}(t)|  \!-\! \frac{t}{2\sqrt{6}}|f_{2,-}(t)| \right] \! \!>\! 0, \nonumber \\
&&  {\rm Im} \,B(t) =   \frac{3 t^2 m_N \sigma_\pi^5 |A_\pi(t) |}{16 (4m_N^2-t)} \nonumber
        \\ && \hspace{2cm} \times  \left[ m_N \sqrt{\frac23}|f_{2,-}(t)|  - |f_{2,+}(t)| \right]  >0, \nonumber
\end{eqnarray}
where the last inequality follows from the data analysis and the Roy-Steiner solutions, yielding
numerically $ m_N \sqrt{2} |f_{2,-}(t)|/ \sqrt{3}  - |f_{2,+}(t)|>0 $.  
Finally, close to the threshold one has 
\begin{eqnarray}
  {\rm Im} \,\Theta (t) & \sim & + (t-4m_\pi^2)^\frac12,  \nonumber \\
   {\rm Im} \,A (t) , {\rm Im} \,J(t),{\rm Im} \,B(t) & \sim & +(t-4m_\pi^2)^\frac52, \label{eq:thresh}
\end{eqnarray}
where the corresponding proportionality constants are {\it positive} and, besides the $\pi\pi\to N \bar N$ amplitudes, also 
involve the pion form factors at the threshold $s= 4 m_\pi^2$. Whereas their
normalization is $A_\pi(0)=1$ and $\Theta_\pi(0)=2m_\pi^2$, from the
meson dominance formulas of our previous work~\cite{RuizArriola:2024udm} one gets
\begin{eqnarray}
  A_\pi(4m_\pi^2) &=& \frac{m_{f_2}^2}{m_{f_2}^2-4 m_\pi^2} \sim 1 + \frac{4m_\pi^2}{m_{f_2}^2}, \nonumber \\
\Theta_\pi(4m_\pi^2) &=& 2 m_\pi^2 + \frac{4m_\pi^2 m_{\sigma}^2}{m_{\sigma}^2 -4 m_\pi^2} \sim 6 m_\pi^2.
\end{eqnarray}

\subsection{Problem of the high-energy completion}

Quite generally, one might hope that the use of the dispersion relations where
${\rm Im }F(s)$ is known up to a sufficiently large value of $s$, such as the 
$N\bar N$ threshold $s_N= 4m_N^2$, allows one to determine accurately
the form factor in the space-like region. Taking for definiteness $D$, one can write
\begin{eqnarray}
  D(-Q^2)= \frac1\pi \int_{4m_\pi^2}^{4 m_N^2} ds \frac{{\rm Im}\, D(s)}{s+Q^2} +
  \frac1\pi \int_{4 m_N^2}^\infty ds \frac{{\rm Im} \, D(s)}{s+Q^2}, \nonumber \\
\end{eqnarray}
where the spectral integral has been divided into a ``better known" part below $s_N$ and a ``less known" part above $s_N$. 
Unitarity provides the absorptive part ${\rm Im} \,D(s)$ up to the $K\bar K$ threshold in terms of the $\pi\pi \to N \bar N$ amplitudes
$f_{+,0}(s)$ and $f_{\pm,2}(s)$ and the pion form factors in the
time-like region, $A_\pi(t)$ and $\Theta_\pi(t)$. However, the high-$s$ part is not known accurately.
Alternatively, one could use subtracted dispersion relations, but then the predictive power is diminished. 
For instance, $D(0)$ would be taken as a low-energy constant, but its value would not be predicted.

The recent work of~\cite{Cao:2024zlf} based on dispersion relations
and incorporating a full Roy-Seiner analysis with all the pertinent
channels, including pions, kaons, and nucleons, has indeed reported a
slow convergence of the dispersive integrals which has some impact
in the space-like region. This is relevant when
confronting with the lattice QCD data~\cite{Hackett:2023rif} for $Q^2
\le 2~{\rm GeV}^2$. To overcome this contingency, additional narrow
resonances have been included in the analysis of~\cite{Cao:2024zlf}.

With this guidance, in the next section we retake the old
idea of the meson dominance to describe the lattice data with a minimal
but realistic amount of information.

\section{Meson dominance approach}

Over the years, the idea of resonance saturation has been extended to all
possible quantum number channels. In our case, we are concerned with the scalar-
and tensor meson dominance.  We first sketch the basic features of the approach and then go
into a detailed analysis, exploring the predictive power of
the meson dominance approach for both GFFs and the mechanical properties of the nucleon.

\subsection{Basic features of meson dominance}

The meson dominance approach, motivated by the large-$N_c$ arguments,
is known to work remarkably well for hadronic form factors at
intermediate space-like momenta (see, e.g.,~\cite{Masjuan:2012sk} and
references therein).\footnote{On the contrary, the determination of the
  time-like behavior of the form factors usually requires abundant data
  in a wide interval and, from the dispersive point of view, an
  extrapolation to the unphysical region. 
  Fortunately, the impact of the time-like details
 on the space-like region considered here is mild.} A form factor $F$ is approximated as,
in general, an infinite sum over narrow resonances in the appropriate
channel,
\begin{eqnarray}
F(-t) = \sum_i \frac{c_i}{m_i^2-t}. \label{eq:fsum}
\end{eqnarray}
Here $m_i$ denotes the resonance mass and $c_i$  its coupling.
Alternatively, the spectral strength is given by 
\begin{eqnarray}
\frac{1}{\pi}{\rm Im}\,F(-t) = \sum_i c_i \delta(m_i^2-t). \label{eq:ssum}
\end{eqnarray}
The short-distance (or large $-t$) constraints (also known as the counting rules) impose relations between $c_i$'s and $m_i$'s. 
In particular, if $\lim_{t\to -\infty} (-t)^n F(t)=0$, one 
obtains the constraints
\begin{eqnarray}
\sum_i c_i m_i^{2k} = 0, \;\;\; k=0, \dots, n-1, \label{eq:co}
\end{eqnarray}
which here are just a manifestation of the super-convergence sum rules~(\ref{eq:srn}).

The previous formulas require a {\it minimal} number of resonances
consistent with normalization and the pQCD sum rules. A natural
extension is to take an infinite number of resonances as required by
large $N_c$ and quark-hadron duality for {\it two-point functions}.
Actually for this case pQCD provides a log-increasing function of
$Q^2$ and quark-hadron duality and consistent with an infinite Regge
spectrum of excited resonances coupled to the current with an
asymptotically constant amplitude. For three-point functions this is
far from trivial since typically in the meson theory one has {\it
  increasing} functions of the logarithm, whereas in pQCD one
obtains {\it decreasing} ones. This was explicitly discussed and
illustrated in Refs.~\cite{RuizArriola:2008sq,RuizArriola:2025wyq,RuizArriola:2025omi} where different schemes are suggested to comply with pQCD.

\subsection{Current-field identities}

In the particular SEM case, 
the current-field identity~\cite{Krolikowski:1967ryy,Raman:1970wq,Raman:1971ur} 
is the simplest way to express the scalar- and tensor-meson dominance,
\begin{eqnarray}
\Theta^{\mu\nu} = \sum_S \frac13 f_S \left( \partial^\mu \partial^\nu - g^{\mu \nu} \partial^2 \right) S +
\sum_T f_T  m_T^2 T^{\mu\nu} \, , \nonumber \\
\end{eqnarray}
where $S$ and $T^{\mu\nu}$ are the scalar $0^{++}$ and tensor $2^{++}$
meson fields, respectively, with $T^{\mu\nu} = T^{\nu\mu}$ and
$T^\mu_\mu=0$.  On-shell, the fields have masses $m_S $ and $m_T$,
respectively, and $\partial^\mu T_{\mu\nu}=0$ (for the complete
Lagrangian see, e.g., \cite{Toublan:1995bk,Ecker:2007us}). Defining
the corresponding sources as $J_S$ and $J_T^{\mu \nu}$,  one gets the
equations of motion,
\begin{eqnarray}
(\partial^2 + m_S^2) S &=&  J_S,  \\  (\partial^2 + m_T^2) T^{\mu\nu} &=& J_T^{\mu\nu} + {\rm dist.} \, ,
\end{eqnarray}
where the distributional contributions are deltas and derivatives
located at $x=0$.  Passing to the momentum space, we get formally (up to
polynomials in $q$)
\begin{eqnarray}
&& \hspace{-5mm} \langle A | \Theta^{\mu\nu} | B \rangle =
\sum_S \frac{f_S}{3} \frac{g^{\mu\nu} q^2-q^\mu q^\nu}{m_S^2-q^2-i\epsilon} \langle A | J_S|B \rangle \nonumber \\ 
&& + \sum_T  f_T \frac{m_T^2} {m_T^2-q^2-i\epsilon} \langle A | \sum_\lambda \epsilon^{\mu\nu}_\lambda
\epsilon^\lambda_{\alpha \beta}J_T^{\alpha \beta}|B \rangle ,
\end{eqnarray}
where $\epsilon^{\mu \nu}_\lambda $ is the spin-2 polarization
tensor, which is symmetric, $\epsilon^{\mu \nu}_\lambda =
\epsilon^{\nu \mu}_\lambda$, traceless, $g_{\mu \nu }\epsilon^{\mu
  \nu}_\lambda=0$, and transverse, $q_\mu \epsilon^{\mu \nu}_\lambda =0$. Clearly, the separate conservation of the scalar and tensor
contributions is manifest according to the Raman decomposition.

\subsection{t-channel unitarity\label{sec:unit}}

From a field theory point of view, the meson dominance formula should
not be taken literally, as it does not incorporate the notion of
subtractions or the pQCD high-momentum behavior. Besides, it is well
known that higher-spin fields have certain problems, particularly due
to the role played by the off-shell behavior of the
propagators~\cite{Toublan:1995bk,Ecker:2007us}. The simplest way to
avoid these issues is to use the meson dominance for the absorptive
parts in the dispersion relations, where by construction the meson
resonances are on the mass shell (see, e.g., ~\cite{Broniowski:2024oyk}
for the pion GFF case). 
Then, one can fix the minimal
number of subtractions needed to satisfy the short-distance
constraints (counting rules) imposed by pQCD.

It is far more practical to compute the absorptive
part of the form factor in the time-like region, where for the graviton 
we have $g \to R \to N \bar N$. At $q^2 \to s+ i \epsilon$ we then have
\begin{eqnarray}
\frac1{\pi} {\rm Im}  \langle N \bar N  | \Theta^{\mu \nu} | 0 \rangle = 
 \sum_R \langle N \bar N | R \rangle  \langle R | \Theta^{\mu \nu} | 0 \rangle \delta(m_R^2-s), \nonumber \\
\end{eqnarray}
and later will reconstruct the dispersive part from the dispersion relation with suitable subtraction constants. 
The vacuum-to-hadron transition amplitudes are
\begin{eqnarray}
  \langle S | \Theta^{\mu \nu} | 0 \rangle = \tfrac{1}{3} f_S  q^2 Q^{\mu\nu}, \;\;\; \langle T | \Theta^{\mu \nu} | 0 \rangle = f_T m_T^2 \epsilon^{\mu \nu}_\lambda, \nonumber \\
\end{eqnarray}
where the extra
factor of ${1}/{3}$ in the scalar case is conventional, chosen such that $\langle S | \Theta| 0 \rangle = f_S q^2$. 
The {\em on-shell} couplings of the resonances to the $N \bar N$ continuum are conventionally taken as \cite{Nagels:1976mc}
\begin{eqnarray}
 && \langle S | N \bar N \rangle = g_{SNN} \, , \\
 && \langle T | N \bar N \rangle = \epsilon^{\alpha \beta}_\lambda \bar v(-p') \left[  g_{TNN}  P^{ \{ \alpha} \gamma^{\beta \} } + 
    f_{TNN}  P^\alpha P^\beta \right] u(p). \nonumber 
\end{eqnarray}
with $v$ denoting  the $\bar N$ spinor. Thus, we get 
\begin{eqnarray}
&& \frac1{\pi }{\rm Im} \langle N  \bar N | \Theta^{\mu\nu} | 0\rangle =
   \sum_S \frac{g_{SNN} f_S}3 \delta(m_S^2-q^2) m_S^2 Q^{\mu\nu}   \nonumber  \\ 
   && \hspace{7mm} +  \sum_{T,\lambda}   \epsilon_\lambda^{\alpha \beta}
P^{ \{ \alpha} \gamma^{\beta \}  }
  \epsilon^{\mu\nu}_\lambda g_{TNN} f_T m_T^2 \delta(m_T^2-q^2) \nonumber \\ && \hspace{7mm} +  \sum_{T,\lambda}   \epsilon_\lambda^{\alpha \beta} P_\alpha P_\beta 
  \epsilon^{\mu\nu}_\lambda f_{TNN} f_T m_T^2 \delta(m_T^2-q^2),
\end{eqnarray}
which naturally complies with a separate conservation for each term, yielding zero when contracted
with $q^{\mu}$. The sum over the tensor polarizations is given by~\cite{Scadron:1968zz,Novozhilov:1975yt}
\begin{eqnarray}
  \sum_{\lambda}   \epsilon_\lambda^{\alpha \beta} \epsilon^{\mu\nu}_\lambda 
= \frac12 \left( Q^{\mu \alpha} Q^{\nu \beta} + Q^{\nu \alpha} Q^{\mu \beta}
  \right) - \frac13 Q^{\mu \nu} Q^{\alpha \beta}. \nonumber \\
   \end{eqnarray}
The on-shell condition $P \cdot q=0$ implies $P_\alpha Q^{\alpha \beta}= P^\beta $
and
$\bar v' \slashed{q} u=0 $ yields $\gamma_\alpha Q^{\alpha \beta}= \gamma^\beta $, 
hence we obtain 
\begin{eqnarray}
&& \sum_{\lambda}   \epsilon_\lambda^{\alpha \beta} P_\alpha P_\beta \epsilon^{\mu\nu}_\lambda 
  = P^\mu P^\nu - \frac13 P^2 Q^{\mu \nu}, \nonumber \\ 
&&
\sum_{\lambda}   \epsilon_\lambda^{\alpha \beta} P^{ \{ \alpha} \gamma^{\beta \} }  \epsilon^{\mu\nu}_\lambda 
= P^{ \{\mu} \gamma^{\nu \} } - \frac13  Q^{\mu \nu} \slashed{P},
\end{eqnarray}
which exactly reproduces the tensor structure in Eq.~(\ref{eq:defwBalt2}).
Therefore, in the narrow resonance,
large-$N_c$ motivated approach we get
\begin{eqnarray}
  \frac1{\pi}  {\rm Im} \,J (s) &=& \frac12     \sum_{T} g_{TNN} f_T m_T^2 \delta( m_T^2-q^2) ,  \nonumber \\
    \frac1{\pi}  {\rm Im} \, B (s) &=& - 2 m_N      \sum_{T} f_{TNN} f_T m_T^2 \delta( m_T^2-q^2) , \nonumber \\
    \frac1{\pi}  {\rm Im} \,\Theta (s) 
    &=&  \sum_S g_{SNN} f_S m_S^2 \delta( m_S^2-q^2), \label{eq:delThet} 
  \end{eqnarray}
and ${\rm Im}\, A(s) = 2 ~{\rm Im}\, J(s) -{\rm Im}\, B(s)$, 
where, as expected, the corresponding spectral strengths get contributions exclusively from the proper spin states: $A$ and $B$ from spin-2 and $\Theta$ from spin-0.

\subsection{Minimal hadronic saturation}

In practice, to avoid proliferation of parameters, one tries to limit the number of mesons (recall footnote~\ref{f1}) to the lowest 
possible number consistent with the constraints.\footnote{For instance, with two mesons included and the $1/(-t)^2$ asymptotic behavior, the condition $c_1 +c_2=0$ cancels the $1/(-t)$ behavior, yielding
$F(t)=F(0) m_1^2 m_2^2/[(m_1^2-t)(m_2^2-t)]$. 
The construction based on Eqs.~(\ref{eq:fsum},\ref{eq:co}) provides a natural interpretation for the widely used multipole fits, for instance the dipole form for the nucleon electromagnetic for factors, since $1/(m^2-t)^2$ is numerically close to $1/[(m_1^2-t)(m_2^2-t)]$ for a suitably chosen $m$ in the considered range of $-t$.}
Note, however, that more precise modeling in the time-like region typically utilizes more resonances and incorporates their width, in an effort 
to reproduce the spectra. An example is the 
Gounaris-Sakurai model~\cite{Gounaris:1968mw,BaBar:2012bdw} for the pion charge form factor, where as many as four $\rho$ resonances are used, including the width effects. 
However, in the space-like region considered here with the aim to understand the lattice QCD data, the meson dominance modeling is not sensitive 
to the details of the time-like region.

In this work we use the asymptotic forms of Eq.~(\ref{eq:asy}) as an inspiration to build the meson saturation model. We do not 
take into account the log corrections, as they would require more precise modeling of the spectral densities, not relevant for the low values of $0<-t< 2 {\rm GeV}^2$, and also would increase the number of free parameters. 
Ignoring the log corrections from $\alpha(t)$ weakens the asymptotic convergence and causes a lack of the sum rule 
(\ref{eq:srn}) with the highest value of $n$ in each channel.\footnote{For the electromagnetic case, 
models including the log corrections, thus satisfying all the super-convergence sum rules, were explored in~\cite{Belushkin:2006qa}.}  As stated above, in a minimum model, one uses the lowest possible number 
of resonances needed to implement the large $-t$ (the short-distance) constraints, as well as the behavior at $t=0$. 
This number is determined by $B(t)$.
Since $B(0)=0$ and at $t \to -\infty$ we have $B(t)\sim 1/(-t)^3$, the lowest number of resonances is four, which leads to the form
\begin{eqnarray}
B(t)=\frac{c_B t}{(1-t/m_{f_2}^2)(1-t/m_{f'_2}^2)(1-t/m_{f''_2}^2)(1-t/m_{f'''_2}^2)}, \nonumber \\ \label{eq:Bmod}
\end{eqnarray}
where $f_2 \dots f'''_2$ are the four lowest-mass $2^{++}$ mesons. 

We assume, accordingly, that the $A$ and $J$ channels couple to the same four resonances, but with different couplings, such that $A(0)=1$, $J(0)=\tfrac12$, and asymptotically Eqs.~(\ref{eq:asy}) hold. Thus, we take\footnote{For a similar construction of the ansatz for the case of the nucleon electromagnetic form factors see, e.g.,~\cite{Masjuan:2012sk}.} 
\begin{eqnarray}
&& A(t)\!=\!\frac{1-c_A t +c_2 t^2}{(1-t/m_{f_2}^2)(1-t/m_{f'_2}^2)(1-t/m_{f''_2}^2)(1-t/m_{f'''_2}^2)}, \nonumber \\
&& J(t)\!=\!\frac{1-c_J t +c_2 t^2}{2(1-t/m_{f_2}^2)(1-t/m_{f'_2}^2)(1-t/m_{f''_2}^2)(1-t/m_{f'''_2}^2)}. \nonumber \\
\label{eq:AJmod}
\end{eqnarray}
Equations~(\ref{eq:J}) and the above formulas are consistent with Eq.~(\ref{eq:Bmod}) with $c_B=c_J-c_A$  (the occurrence of the same $c_2$ 
in both formulas (\ref{eq:AJmod}) follows from $B(t) \sim 1/t^3$).

In the scalar channel, where $\Theta(0)=m_N$, we take for simplicity two resonances,
\begin{eqnarray}
\Theta(t)=\frac{m_N}{(1-t/m_{\sigma}^2)(1-t/m_{f_0}^2)}, \label{eq:Thmod}
\end{eqnarray}
where the power agrees with the counting rules for the corresponding pQCD kernel. 
Here $m_\sigma$ is 
an effective mass of the sigma meson and $f_0$ is the scalar resonance $f_0(980)$. Again, more precise modeling would require more realistic and complicated spectral functions, which is expected to be irrelevant in the space-like momentum region explored in lattice QCD simulations. 

Naturally, Eqs.~(\ref{eq:AJmod},\ref{eq:Thmod}) can be written, via pole expansion of a meromorphic function, as sums of simple poles of the form~(\ref{eq:fsum}), where the residues ``conspire'' in such a way that the short-distance constraints are satisfied.  
Some properties of the fitted functions are presented in Appendix~\ref{app:fit}.

\subsection{Fitting procedure \label{sec:fit}}

We treat $m_\sigma$, $c_A$, $c_J$, and $c_2$ as free model parameters. The mass of $f_0$ is fixed at $m_{f_0}=980$~MeV, while 
for the tensor masses we probe two sets, largely differing in the second and third excited states (all masses in GeV),
\begin{eqnarray}
{\rm I}\!: &&~m_{f_2}\!= \! 1.275, m_{f'_2}\!=\!  1.517, m_{f''_2}\!=  \!1.565, m_{f'''_2}\!=\!1.936, \nonumber \\
{\rm II}\!: && ~m_{f_2}\!=\!  1.275, m_{f'_2}\!= \! 1.430, m_{f''_2}\!=\!  1.517, m_{f'''_2}\!=\!1.565, \nonumber \\
\label{eq:sets}
\end{eqnarray}
with set I corresponding to the first four entries in the PDG (Particle Data Group)~\cite{ParticleDataGroup:2024cfk} summary tables, and set II to the PDG listings. Comparing the fits made with sets I and II allows for an estimate of a ``systematic'' error in our analysis. It turns out that in the range of the space-like lattice data, this error is much smaller (relatively, a few percent)
from the statistical error originating from the uncertainty in the data. For this reason we combine the systematic and statistical errors in the results presented below. 

Since the lattice data correspond to $m_\pi=170$~MeV, we use 
correspondingly an increased nucleon mass, $m_N=970$~MeV~\cite{Alvarez-Ruso:2013fza}. We do not change the values of the resonance masses, as they are expected to be at the level of a few percent. Accordingly, 
our results for the nucleon GFFs correspond to $m_\pi=170$~MeV. 
Based on our experience for the pion~\cite{Broniowski:2024oyk}, we anticipate that the results of 
the chiral evolution to the physical point $m_\pi=140$~MeV are not very significant, at a level of a few percent for the values of the model parameters.

\subsection{Results}

The joint fit to the available MIT lattice data for $A$, $J$, and $D$ (for which we use Eq.~(\ref{eq:Drel})) yields 
for set I
\begin{eqnarray}
&& c_A=0.62(5)~{\rm GeV}^{-2}, \;\;c_J=0.87(6)~{\rm GeV}^{-2}, \nonumber \\
&& c_2=0.15(5)~{\rm GeV}^{-4}, \;\;m_\sigma=0.64(4)~{\rm GeV},
\end{eqnarray}
and for set II 
\begin{eqnarray}
&& c_A=0.83(6)~{\rm GeV}^{-2}, \;\;c_J=1.12(7)~{\rm GeV}^{-2}, \nonumber \\
&& c_2=0.25(5)~{\rm GeV}^{-4}, \;\;m_\sigma=0.64(4)~{\rm GeV},
\end{eqnarray}
where the errors are statistical. The change of the $c_i$ coefficients between the two sets follows the features explained in Appendix~\ref{app:fit}.
For both sets the values of $\chi^2/{\rm DOF}$ are $\sim 0.4$, indicating that the nature of the lattice
data errors involves systematic uncertainty and/or correlations. 
In our fits, the values of $c_A$, $c_J$, and $c_2$ change to compensate for the effect of different masses of the $m_{f_2}$ states in parametrizations I and II in the range of the data. The value of $m_\sigma$ is in agreement within the uncertainties with the fit for the pion case in~\cite{Broniowski:2024oyk}, where we obtained $m_\sigma=0.65(3)$~GeV at $m_\pi=170$~MeV.
 
The correlation matrix, for the rows and columns in the sequence of parameters given above, is, for case I,
\begin{eqnarray}
\rho=\left(
\begin{array}{cccc}
 1 & 0.7 & -0.9 & 0.3 \\
 0.7 & 1 & -0.8 & 0.1 \\
 -0.9 & -0.8 & 1 & -0.1 \\
 0.3 & 0.1 & -0.1 & 1 \\
\end{array}
\right)
\end{eqnarray}
(for case II, it is similar).
We note that $m_\sigma$ is weakly correlated to the other parameters, as this correlation occurs only by their common appearance in $D$. On the other hand, the $c_i$ coefficients are strongly correlated. The correlations are taken into account in the error propagation to the form factors, shown in 
the figures below with the bands, indicating the 68\% confidence levels. 

\begin{figure}[t]
    \centering
    \includegraphics[width=0.45\textwidth]{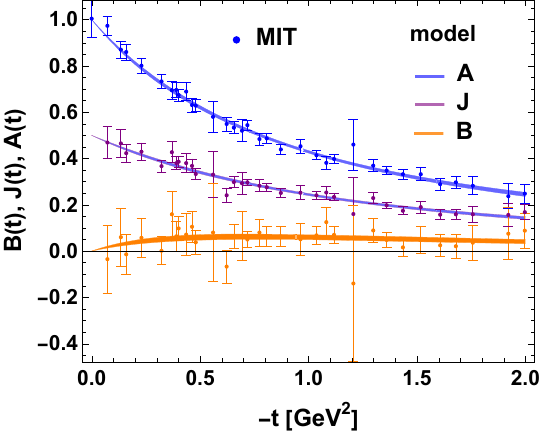}
    \caption{Gravitational form factors of the nucleon, $A$, $J$, and $B$, plotted as functions of the space-like momentum transfer $-t$. The MIT lattice QCD data for $A$ and $J$ are taken from~\cite{Hackett:2023rif}, while those for $B$ follow from the relation $B=2J-A$ with errors added in quadrature. The meson dominance model fit is shown with the error bands, whose widths reflect the uncertainty in the parameters (see the text for details). The MIT data and the fits are for $m_\pi=170$~MeV. \label{fig:JAB}}
\end{figure}

\begin{figure}[t]
    \centering
    \includegraphics[width=0.45\textwidth]{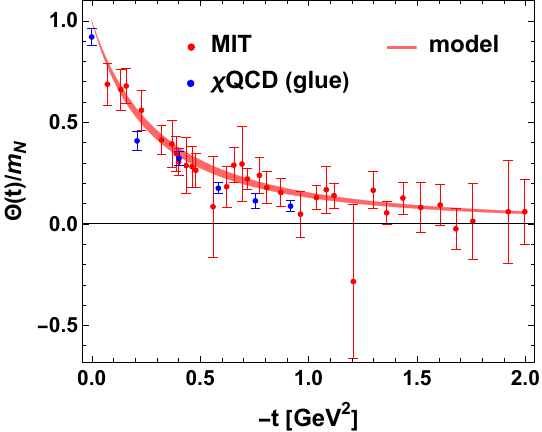}
  \caption{Scalar (trace anomaly) gravitational form factor $\Theta$ of the nucleon, divided with the nucleon mass $m_N$ and plotted as a function of the space-like momentum transfer $-t$. The red data points labeled MIT are obtained from the data~\cite{Hackett:2023rif} via the relation~(\ref{eq:Th}), with errors added in quadrature.  The meson dominance model fit is shown with the error band, whose width reflects the uncertainty of the effective $\sigma$ mass (see text for details). The MIT data and the fit are for $m_\pi=170$~MeV. 
We also show (blue points), for a rough comparison, the glue contribution to the trace anomaly form factor at $m_\pi=253$~MeV from the 
$\chi$QCD Collaboration~\cite{Wang:2024lrm}. \label{fig:Th}}
\end{figure}

In Fig.~\ref{fig:JAB} we plot the form factors $A$, $J$, and $B$. We get a proper description of the MIT lattice data~\cite{Hackett:2023rif}. The bands in the figures indicate the combined systematic error from the two choices of the tensor meson spectrum~(\ref{eq:sets}), and from the statistical error reflecting the uncertainties in the fitted lattice data. We note that the systematic error for $A$ and $J$, estimated by comparing sets I and II, is significantly smaller from the statistical error (a few percent), reflecting the fact that at the available range of $-t$ one cannot resolve the details of the model.

For the case of $B$, the values for the data were obtained via the relation~(\ref{eq:J}), with the errors added in quadrature. We note the smallness of $B$, yet it is significantly larger from zero at $-t>0$ compared to the error bars.
 The LHPC Collaboration~\cite{LHPC:2007blg} computed the quark parts of the GFF's and found a similar trend. Interestingly, the observed smallness was predicted long ago by Renner~\cite{Renner:1970sbf} within the quark model. The feature is reproduced in the Chiral Soliton
model~\cite{Goeke:2007fq} and less accurately in the Skyrme model~\cite{Cebulla:2007ei}.

The scalar GFF $\Theta(t)$ is plotted in Fig.~\ref{fig:Th}, where the data were obtained from~\cite{Hackett:2023rif} via the combination of Eq.~(\ref{eq:Th}) with errors added in quadrature. The minimum meson dominance formula~(\ref{eq:Thmod}) fits the data properly, and as already remarked, with $m_\sigma$ compatible within uncertainty to the case of the pion GFFs~\cite{Broniowski:2024oyk}. The present lattice errors are too large and the range of $-t$ too small to test  more precisely the asymptotic behavior of Eq.~(\ref{eq:thasy}).

\begin{figure}[t]
    \centering
    \includegraphics[width=0.45\textwidth]{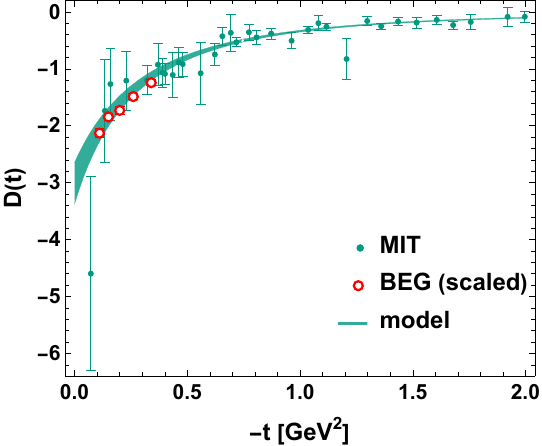}
  \caption{Gravitational form factor $D$ of the nucleon, plotted as a function of the space-like momentum transfer $-t$. The MIT lattice QCD data are from~\cite{Hackett:2023rif}.  The meson dominance model fit is shown with an error band whose width reflects the uncertainty in the parameters 
  (see the text for details). The MIT lattice data and the fit are for $m_\pi=170$~MeV. The points labeled BEG (open circles) are from DVCS obtained for the quark part in~\cite{Burkert:2018bqq} and rescaled as described in the text. \label{fig:D}}
\end{figure}

Figure~\ref{fig:D} shows the form factor $D$.  In this case the model description follows from Eq.~(\ref{eq:Drel}). 
Again, we can see a proper description of the MIT lattice data, with the error band getting broader as $-t$ is decreased. 
For comparison, we also show the physical data from deeply virtual Compton scattering (DVCS) on the proton measured at CLAS~\cite{CLAS:2015uuo} and extracted in~\cite{Burkert:2018bqq} for the 
quark part of $D(t)$. The data points were obtained from Fig.~11 of~\cite{Burkert:2023wzr} by utilizing the 
relation $D_{u+d}=\tfrac{18}{25}{\cal C_H}(t)$, cf. Eq.~(40) in~\cite{Burkert:2023wzr}. To compare to $D(t)$ including the gluons, we 
assume that the form factors for the quarks and gluons have the same shape in $-t$,\footnote{This in general may not be the case, as the 
form factors $\bar{c}_a(t)$, dependent on the 
renormalization scheme and scale, may be different for various parton species. However, in models where 
all the gluons are radiatively generated by the QCD evolution from a low quark model scale, the shapes of the gluon and quark form factors are equal~\cite{Broniowski:2007si}.} which allows 
for the scaling $D(t) = (\langle x\rangle_q+\langle x\rangle_g)/ \langle x\rangle_q)D_q(x)$. We take the world average $\langle x\rangle_g\simeq 0.41$ 
at $\mu=2$~GeV, using the compilation of Fig.~8 in~\cite{Fan:2022qve}. 
Thus extracted DVCS data lie within the band corresponding to our fit to the MIT data. 

Note that in our model, at space-like momenta, $A$, $J$, $B$, and $\Theta$ are positive-definite, whereas $D$ is negative-definite. Hence all form factors have a definite sign. Except for $\Theta$, which according to Eq.~(\ref{eq:thasy}) should change the sign, the obtained sign 
behavior is compatible with the 
asymptotics of Eq.~(\ref{eq:asy}).

\subsection{$D$-term \label{sec:dterm}}

The obtained value of the $D$-term corresponding to the plot in Fig.~\ref{fig:D} is
\begin{eqnarray}
D(0) = - 3.0(4).
\end{eqnarray}
The mechanical radius, as defined in~\cite{Polyakov:2018zvc}, is
\begin{eqnarray}
\langle r^2 \rangle_{\rm mech} = \frac{6 D(0)}{\int_0^\infty d(-t) D(t)} = [0.72(5)~{\rm fm}]^2, 
\end{eqnarray}
which is a value comparable to other estimates (see~\cite{Goharipour:2025yxm} for a recent review).

\section{Transverse densities \label{sec:td}}

\subsection{Motivation}

While not directly accessible in experiment, the transverse
two-dimensional densities of hadrons, evaluated in the infinite
momentum frame, were promoted
in~\cite{PhysRevD.15.1141,Burkardt:2000za,Diehl:2002he,Burkardt:2002hr,Miller:2010nz,Freese:2022fat}
as the proper spatial distributions, dependent only on the intrinsic
features of the probed state. The main advantage is a possibility, in
the case where positivity holds, of a probabilistic interpretation of
the parton distributions. In particular, the transverse density of
quarks of a particular flavor is manifestly positive
definite~\cite{Burkardt:2002hr,Pobylitsa:2002iu,Diehl:2002he}. Analogously,
the transverse $\Theta^{++}$ distribution is also positive definite
(in the light-cone gauge) for any hadronic state, as shown generally
in~\cite{Broniowski:2024mpw}, 
\begin{eqnarray}
 \int dx^- \Theta^{++} (b,x^-)|_{x^+=0} > 0.
\end{eqnarray} 
Moreover, the transverse densities are related to the transverse mechanical
properties, as elaborated below, which is of attractive
interpretational merit.

\subsection{Definitions}

We use the light-cone coordinate system in the convention $a^\pm=1/\sqrt{2} (a^0 \pm a^3)$. 
The infinite-momentum Breit frame (IMBF), also known as the Drell-Yan frame~\cite{Drell:1969km}, is defined as 
\begin{eqnarray}
&& P^1=P^2=0, \;\; P^-=\frac{m_N^2-\tfrac{1}{4}t}{2P^+}, \nonumber \\
&& q^+=q^-=0, 
\end{eqnarray}
with $P^+=p^+=p'^+\to \infty$. In this frame, the Dirac spinors are
\begin{eqnarray}
u(p,s) =\sqrt{p^+}\begin{pmatrix}1\\ \sigma_3 \end{pmatrix} \chi_s,
\end{eqnarray}
with $\chi_s$ denoting a two-component Pauli spinor and $\bar{u}'(p',s') u(p,s) = 2P^+ \delta_{ss'}$. 
IMBF is used in the formulas of this and the following subsections.

The matrix elements in the transverse coordinate space are defined via the two-dimensional Fourier transforms of the basic SEM matrix 
elements, namely,
\begin{eqnarray}
\hspace{-7mm} T^{\mu \nu}({b})= \int \frac{d^2 q_\perp}{(2\pi)^2 2P^+} e^{-i {\bf b} \cdot {\bf q}_\perp} \langle p^\prime,s | \Theta^{\mu\nu}(0) | p,s\rangle.
\end{eqnarray}
For the $++$ component one finds
\begin{eqnarray}
\hspace{-7mm} T^{++}({b})= \frac{P^{+2}}{m_N}\int \frac{d^2 q_\perp}{(2\pi)^2} e^{-i {\bf b} \cdot {\bf q}_\perp} A(-q_\perp^2), \label{eq:T++}
\end{eqnarray}
hence it depends only on the $A$ form factor. We may thus define 
\begin{eqnarray}
&& A(b)= \int \frac{d^2 q_\perp}{(2\pi)^2} e^{-i {\bf b} \cdot {\bf q}_\perp} A(-q_\perp^2) = \nonumber \\
&& \hspace{11mm} \int_0^\infty \frac{q_\perp d q_\perp}{2\pi} J_0(b\,q_\perp) A(-q_\perp^2), \label{eq:Ab}
\end{eqnarray}
with an obvious normalization $\int_0^\infty 2\pi b\,db\,A(b)=1$. According to Eq.~(\ref{eq:T++}), $A(b)$ can 
be interpreted as the distribution of $P^{+2}/m_N$ in the transverse coordinate plane. 

For the trace anomaly form factor we obtain
\begin{eqnarray}
\hspace{-7mm} \Theta(b)\equiv 2T^{+-}({b})-T^{11}({b})-T^{22}({b})= \nonumber \\
\int \frac{d^2 q_\perp}{(2\pi)^2} e^{-i {\bf b} \cdot {\bf q}_\perp} \Theta(-q_\perp^2), \label{eq:Tth}
\end{eqnarray}
with $\int_0^\infty 2\pi b\,db\,\Theta(b)=m_N$. 

The transverse-coordinate spin density along the $z$-axis, $J(b)$, is evaluated as (cf. Eq.~(76) of~\cite{Lorce:2017wkb} and the discussion 
of the associated symmetric prescription for $\Theta^{\mu \nu}$)
\begin{eqnarray}
&& J(b)=\int \frac{d^2q_\perp}{(2\pi)^2} e^{-i \bm{q_\perp} \cdot \bm{b}} \left [ J(-q_\perp^2) +q_\perp^2 \frac{d J(-q_\perp^2)}{dq_\perp^2} \right ] = \nonumber \\
&& \hspace{7mm} \frac{b}{4\pi} \int_0^\infty dq_\perp J_1(bq_\perp)  \left [ q_\perp^2  J(-q_\perp^2) \right ], \label{eq:Jb}
\end{eqnarray} 
where in  getting to  the last line we have integrated by parts and used the relation for the Bessel functions, $\frac{d}{dz}J_0(z)=-J_1(z)$.

The transverse density of $D$ is related to the transverse (mechanical) components of SEM, discussed in detail in Sec.~\ref{sec:mech}.

\subsection{Dispersive sum rules and the $b\to 0$ limit \label{sec:disp}}

For the transverse density defined via
\begin{eqnarray}
&& F(b)=\int \frac{d^2 q_\perp}{(2\pi)^2} e^{-i {\bf b} \cdot {\bf q}_\perp} F(-q_\perp^2) = \nonumber \\
&& \hspace{1cm} \frac{1}{2\pi}\int q_\perp\,dq_\perp J_0(b q_\perp) F(-q_\perp^2) , \label{eq:Fgen}
\end{eqnarray}
the dispersion relation yields, after carrying out the $q_\perp$ integration (which can be interchanged with the integration over $s$ for convergent 
integrals), the following formula~\cite{Miller:2010tz}:
\begin{eqnarray}
F(b) = \frac{1}{2\pi^2} \int_{4m_\pi^2}^\infty ds K_0(b \sqrt{s}) \, {\rm Im}\, F(s). \label{eq:FbK}
\end{eqnarray}
For the special case of the $J$ form factor, with the definition (\ref{eq:Jb}), one finds instead
\begin{eqnarray}
J(b) = \frac{b}{4\pi^2} \int_{4m_\pi^2}^\infty ds \sqrt{s} K_1(b \sqrt{s}) \, {\rm Im}\, J(s). \label{eq:JbK}
\end{eqnarray}
Because at asymptotically large arguments the Bessel functions $K_n(z)\sim e^{-z} \sqrt{\pi/(2z)}$, integrals in (\ref{eq:FbK},\ref{eq:JbK}) are
convergent.
 
At $b \to 0$, the integration kernel in  (\ref{eq:FbK}) is singular, since $K_0(b \sqrt{s}) \sim - \tfrac{1}{2} \log(b^2 s)+{\rm const}+{\cal O}(b^2 \log b)$. 
However, the $\log(b^2)$ term and the constant are canceled via the super-convergence sum rule (\ref{eq:srn}) for $n=1$. This yields a model independent 
general sum rule
\begin{eqnarray}
F(b=0) = -\frac{1}{4\pi^2} \int_{4m_\pi^2}^\infty ds \log s \, {\rm Im}\, F(s), \label{eq:F0}
\end{eqnarray}
holding for $F=A$, $B$, $\Theta$, and $D$. Equation (\ref{eq:F0}) may be viewed as a spectral sum rule, which due to the $-\log s$ weight\footnote{One does not need to keep a dimensionful scale here, since $\log(s/\mu^2)=\log s - \log \mu^2$, and the second term is canceled thanks to the sum rule (\ref{eq:srn}).} collects contributions from the whole physical region in $s$, and not as one might naively expect, 
predominantly from the low-$s$ values. At lower $s$, this weight is positive, and at large $s$ it is negative, but since the spectral strength has no definite sign as implied by the super-convergence sum rule, it is not possible to make a priori statements concerning the sign of $F(b=0)$. 

Moreover, there is an important methodological warning following from
Eq.~(\ref{eq:F0}) which we have not seen discussed in the
literature.  The formula indicates a strong sensitivity also to the
high-$s$ part of the spectrum, which at present is largely unknown and
dependent on the particular model used.  Another way to see this more
directly is from Eq.~(\ref{eq:Fgen}) and noticing that the tails of
the form factors in Figs.~\ref{fig:JAB}-\ref{fig:D} extending outside
of the range of the data, and still much below the in principle known
pQCD asymptotics, are strong.  This uncertainly must be kept in mind
when obtaining the transverse densities from models or fits to data,
including the mechanical properties discussed in Sec.~\ref{sec:mech}.

For the curvature at $b=0$, an analogous reasoning to the one leading to Eq.~(\ref{eq:F0}) can be carried out with the sum rule (\ref{eq:srn}) for $n=2$, which results in 
\begin{eqnarray}
&&F(b) = F(b=0) -\frac{b^2}{16\pi^2} \int_{4m_\pi^2}^\infty ds \,s\log s \, {\rm Im}\, F(s) \nonumber \\ 
&&\hspace{4cm} +{\cal O}{(b^4 \log b)}. \nonumber \\ \label{eq:F2}
\end{eqnarray}
According to the discussion in Sec.~\ref{sec:pQCD}, with proper pQCD constraints, it holds for $F=A$, $B$, and $D$.
Similarly, for the $b^4$ contribution we obtain
\begin{eqnarray}
&&F(b) = F(b=0)+ F'(b=0)b^2 \nonumber \\
&& \hspace{2mm}-\frac{b^4}{256\pi^2} \int_{4m_\pi^2}^\infty ds \,s^2\log s \, {\rm Im}\, F(s) +{\cal O}{(b^6 \log b)},  \label{eq:F4}
\end{eqnarray}
valid for $F=B$ and $D$, where the sum rule (\ref{eq:srn}) for $n=3$ holds.
The higher order terms in $b$ contain the also $\log b$ pieces.

For $J$, using $K_1(b \sqrt{s}) \sim 1/(b \sqrt{s}) + b \sqrt{s} \tfrac{1}{4} \log(b^2 s)+{\rm const.}+{\cal O}(b^2 \log b)$, we find
\begin{eqnarray}
\hspace{-5mm} J(b) = \frac{b^2}{16\pi^2} \int_{4m_\pi^2}^\infty ds \,s\log s \, {\rm Im}\, J(s) +{\cal O}{(b^4 \log b)}. \label{eq:J2}
\end{eqnarray}

Equations (\ref{eq:F2}-\ref{eq:J2}) are even more sensitive to the
high-lying part of the spectrum than Eq.~(\ref{eq:F0}) due to extra
powers of $s$ in the integrand, hence the uncertainty discussed above
is enhanced.

The regular behavior of $F(b=0)$ from Eq. (\ref{eq:F0}) is in contrast to the case 
of the pion GFFs, where the long $q_\perp$ tails 
(pQCD predicts $A^\pi(t)\sim D^\pi(t) \sim \alpha(t)/(-t)$ and $\Theta^\pi \sim \alpha(t)^2$~\cite{Tong:2021ctu}) make the integrals in the Fourier 
transforms only conditionally convergent and thus the transverse densities of $A$ and $\Theta$ for the pion become singular at $b=0$~\cite{Miller:2009qu,Broniowski:2024mpw} and of a definite sign. The appearance of this singularity is clear from the present discussion. The pion gravitational form factors $A$ and $D$, as well as the pion electromagnetic form factor, have the 
corresponding spectral densities behaving as $\sim 1/(s \log^2 s)$. Then the integrand of~(\ref{eq:F0}) is asymptotically $1/(s \log s)$, which leads to divergence at $b\to 0$.

We conclude the general considerations of this subsection by remarking that for the nucleon electromagnetic form factor, $F_1(t) \sim \alpha^2/(-t)^2$, Eqs.~(\ref{eq:F0},\ref{eq:F2}) are valid, while for 
$F_2(t) \sim \alpha^2/(-t)^3$, Eq.~(\ref{eq:F4}) also applies.

\subsection{Large-$b$ asymptotics}

Using Eqs.~(\ref{eq:thresh}) for the threshold behavior of the
spectral densities at the $2\pi$ branch point and formulas
(\ref{eq:FbK}-\ref{eq:JbK}), we arrive at the following behavior at
$b\to \infty$:
\begin{eqnarray}
&& A(b), B(b) \sim + e^{-2m_\pi b}/b^4, \nonumber \\
&& J(b)\sim + e^{-2m_\pi b}/b^3,  \nonumber \\
&& \Theta(b)\sim + e^{-2m_\pi b}/b^2, \label{eq:basy}
\end{eqnarray}
where $+$ indicates a positive constant. The powers of $b$ are of a
kinematic origin, related to the threshold behavior. The common
positive sign stems from a combination of unitarity and analyticity
driven by the data analysis (see discussion in Sec.~\ref{sec:spec-low}).
The powers of $b$ are the same as for the pion (cf. Eq.~(63) of~\cite{Broniowski:2024mpw}), which follows from the formalism 
brought up in Sec.~\ref{sec:spec-low}.
The P-wave case of the electromagnetic form factor of the pion, using the time-like experimental data, was analyzed in~\cite{RuizArriola:2025wyq}.

\subsection{Meson dominance results}

\begin{figure}[t]
    \centering
    \includegraphics[width=0.45\textwidth]{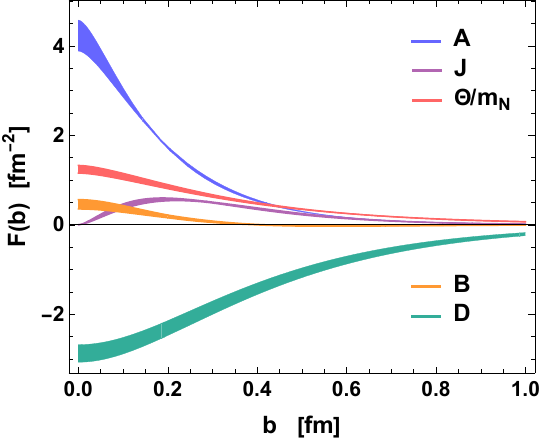}
    \caption{Transverse densities of $A$, $J$, $\Theta/m_N$, $B$, and $D$, plotted as functions of the transverse coordinate $b$, following from the meson dominance model fitted to the MIT lattice data at $m_\pi=170$~MeV. \label{fig:tranden} }
\end{figure}

We now return to our meson dominance model.
Inserting the spectral strength of the form~(\ref{eq:ssum}) into~(\ref{eq:FbK}) or~(\ref{eq:JbK})  we find simple expressions involving sums of the Bessel functions  $K_n(b m_i)$. The results for the basic form factors are presented in 
Fig.~\ref{fig:tranden}.
For all cases, $\int 2\pi b\,db\,F(b)=F(t=0)$.
The bands in the figures of this subsection indicate the errors discussed in Subsection~\ref{sec:fit}, but do not include the model uncertainties
discussed in Sec.~\ref{sec:disp}, as they are difficult to assess.

In~\cite{Broniowski:2024mpw}, within a discussion for the pion, we
have shown that $A(b)$ is positive definite at an operator level,
i.e., independent of the hadron state in which the matrix element is
taken.  The present model for $A$ complies with that general result. We
note that in our model, the transverse density $A(b)$ is positive
definite, as required by the positivity condition.

\begin{figure}[t]
    \centering
    \includegraphics[width=0.45\textwidth]{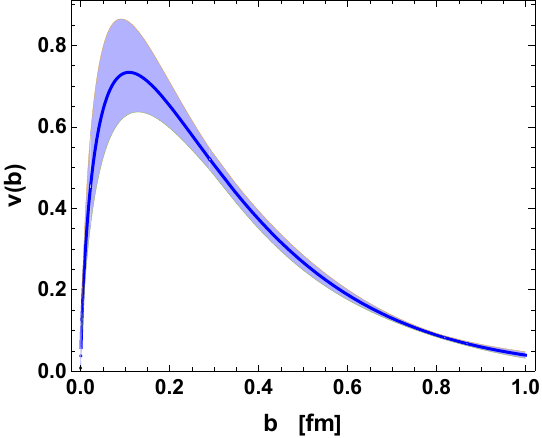}
    \caption{Tangential velocity profile, defined in analogy to classical mechanics in Eq.~(\ref{eq:v}), plotted as a function of the transverse coordinate $b$. \label{fig:vel}}
\end{figure}

Classically, the angular momentum is ${\bf J}={\bf r} \times {\bf p} = m \, {\bf r} \times {\bf v}$. Following this analogy, we define the tangential velocity 
profile in the nucleon as
\begin{eqnarray}
v(b)=\frac{J(b)}{b\, \Theta(b)}. \label{eq:v}
\end{eqnarray}
As is seen from Fig.~\ref{fig:vel}, reflecting the behavior of the involved transverse densities, $v(b)$ grows linearly from 0 at low $b$, reaches a maximum around $0.1$~fm, and then decreases approaching 0.

\section{Transverse mechanical properties \label{sec:mech}}

\subsection{Definitions and properties \label{sec:trp}}

The formulas displayed in this Section mirror the derivations by Polyakov and Schweitzer~\cite{Polyakov:2018zvc}, but for the two-dimensional transverse densities~\cite{Panteleeva:2021iip}. 
The transverse ($i,j=1,2$) part of SEM is
\begin{eqnarray}
&& \hspace{-7mm} T^{ij}({b})= \delta^{ij} p(b) + \left ( \frac{b^i b^j}{b^2} - \frac{1}{2} \delta^{ij} \right ) s(b) = \label{eq:Tij} \\
&& \frac{1}{4m_N}\int \frac{d^2 q_\perp}{(2\pi)^2} e^{-i {\bf b} \cdot {\bf q}_\perp} 
( q_\perp^i q_\perp^j -\delta_{ij}q_\perp^2)D(-q_\perp^2). \nonumber
\end{eqnarray}
The transverse pressure is therefore given by
\begin{eqnarray}
\hspace{-4mm} p(b)=-\frac{1}{16\pi m_N} \int_0^\infty q_\perp\,dq_\perp J_0(q_\perp b) q_\perp^2 D(-q_\perp^2), \label{eq:p1}
\end{eqnarray}
whereas for the stress we get 
\begin{eqnarray}
s(b)=-\frac{1}{8\pi m_N} \int_0^\infty q_\perp\,dq_\perp J_2(q_\perp b) q_\perp^2 D(-q_\perp^2). \label{eq:s1}
\end{eqnarray}
Since $J_0(z)=1-z^2/4+\dots$, at the origin the pressure is positive and concave, as long as $D(-q_\perp^2)$ is negative. 
Similarly, $J_2(z)=z^2/8+\dots$ implies vanishing and convex stress near $b=0$.

From Eq.~(\ref{eq:p1}) it is obvious that 
\begin{eqnarray}
\int_0^\infty db \, 2\pi b\,p(b)=0, \label{eq:zp}
\end{eqnarray}
as required by a classical analog of the stability condition. One can also show that 
\begin{eqnarray}
\int_0^\infty db \, 2\pi b\,s(b)=-\frac{1}{2m_N} \int_0^\infty q_\perp  dq_\perp D(-q_\perp^2).
\end{eqnarray}

From the conservation of a static SEM, $\partial/\partial b^i \, T^{ij}=0$, the equation
\begin{eqnarray}
p'(b)+\frac{1}{2} s'(b) + \frac{s(b)}{b}=0 \label{eq:de}
\end{eqnarray}
follows. As a check, this relation can be verified with the identity between the Bessel functions, $J'_0(z)+J'_2(z) + 2J_2(z)/z =0$, applied to Eqs.~(\ref{eq:p1}) and (\ref{eq:s1}).
Integration of (\ref{eq:de}) by parts 
yields 
\begin{eqnarray}
p(0)=\int_0^\infty \frac{s(b)}{b} db.
\end{eqnarray}
Other relations can be derived, stemming from the fact that $p(b)$ and
$s(b)$ are defined via integrals with common function $D(-q_\perp^2)$
and the two Bessel functions:
\begin{eqnarray}
&& \int_0^\infty p(b)db=\frac{1}{2} \int_0^\infty s(b)db, \nonumber \\
&& \int_0^\infty b^2 p(b)db=-\frac{1}{6} \int_0^\infty b^2 s(b)db, \nonumber \\
&& \dots
\end{eqnarray}
In particular, the following relation with the $D$-term holds: 
\begin{eqnarray}
D(0) = 2 m_N \int_0^\infty \!\!\!\!\! 2 \pi b\,db\,  b^2 p(b) =  - \frac{m_N}{2} \int_0^\infty \!\!\!\!\! 2 \pi b\,db\, b^2 s(b). \nonumber \\
\end{eqnarray}

The spectral representation is
\begin{eqnarray}
&& p(b) = \frac{1}{16\pi^2 m_N} \int_{4m_\pi^2}^\infty ds K_0(b \sqrt{s}) s\, {\rm Im}\, D(s), \nonumber \\
&& s(b) = -\frac{1}{8\pi^2 m_N} \int_{4m_\pi^2}^\infty ds K_2(b \sqrt{s}) s\, {\rm Im}\, D(s), \label{eq:pbK}
\end{eqnarray}
from where, in conjunction with the super-convergence sum rules, it follows that 
\begin{eqnarray}
&& p(b=0) = -\frac{1}{32\pi^2 m_N} \int_{4m_\pi^2}^\infty ds \, s \log s \, {\rm Im}\, D(s),  \label{eq:p0}  \\
&& \left . \frac{ds(b^2)}{db^2} \right |_{b=0}= \frac{1}{64\pi^2 m_N} \int_{4m_\pi^2}^\infty ds \, s^2 \log s \, {\rm Im}\, D(s). \nonumber
\end{eqnarray}
Note the sensitivity to the high-$s$ time-like region. 

The asymptotic behavior, reflecting Eq.~(\ref{eq:delThet}) and obtained straightforwardly via Eq.~(\ref{eq:pbK}), is 
\begin{eqnarray}
&& p(b) \sim - e^{-2m_\pi b}/b^2, \nonumber \\
&& s(b) \sim + e^{-2m_\pi b}/b^2.
\end{eqnarray}
Note that it is governed by the behavior of ${\rm Im}\,\Theta(s)$, which is dominant at the $2\pi$ threshold.

Finally, using the fact that $2 T^{+-}(b) = \epsilon(b) $ is the transverse energy density, one has the natural relation
\begin{eqnarray}
  \theta(b) = \epsilon(b) - 2 p(b) 
\end{eqnarray}
Note that we intuitively expect $\epsilon(b) >0$, although $p(b)$
changes sign and $\theta(b)$ need not be positive (it was not in the
pion case~\cite{Broniowski:2024mpw}).  We have checked that indeed $\epsilon(b)$ is positive in the present meson dominance analysis.

\subsection{Meson dominance results}

\begin{figure}[t]
    \centering
    \includegraphics[width=0.45\textwidth]{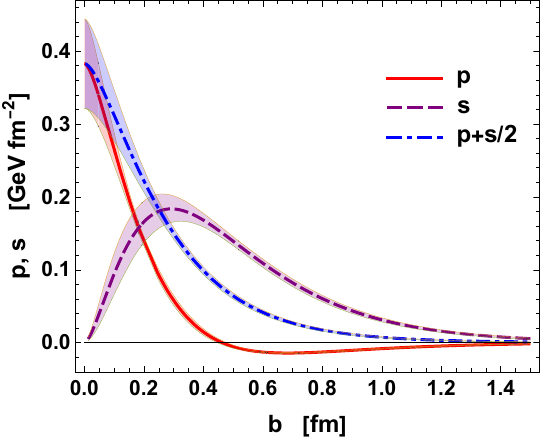}
    \caption{Pressure $p$ and stress $s$ in the nucleon, evaluated in the meson dominance model fitted to the MIT lattice data at $m_\pi=170$~MeV. The combination $p + \tfrac{1}{2}s$ is positive, as required by stability. \label{fig:ps}}
\end{figure}

\begin{figure}[t]
    \centering
    \includegraphics[width=0.45\textwidth]{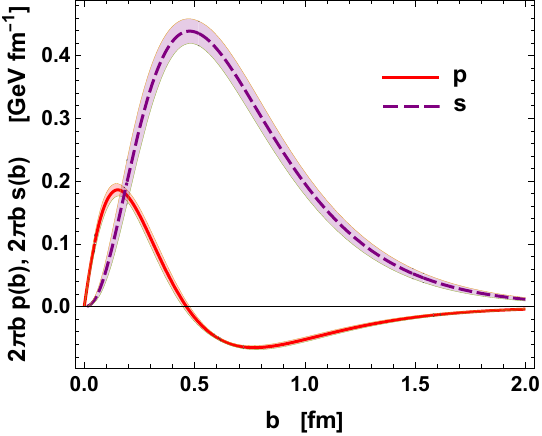}
    \caption{Pressure $p$ and stress $s$ in the nucleon in the meson dominance model, multiplied with $2\pi b$.  
    The pressure changes sign from $+$ to $-$ as $b$ increases, and $\int_0^\infty 2\pi b p(b) db=0$, as required by stability. \label{fig:psb}}
\end{figure}

The transverse pressure and stress evaluated in the meson dominance model of the previous sections are plotted in Fig.~\ref{fig:ps}.
We also show the combination $p(b)+s(b)/2$, whose positivity is the local criterion for classical stability~\cite{Polyakov:2018zvc} for the two-dimensional case. 
We see from the figure that this criterion is satisfied.  In Fig.~\ref{fig:psb} we display the pressure and stress multiplied with $2\pi b$. It is 
clearly seen that the positive pressure inside the nucleon is compensated by the negative pressure outside, in agreement with Eq.~(\ref{eq:zp})
and expectations from classical stability conditions.

\begin{figure}[t]
    \centering
    \includegraphics[width=0.473\textwidth]{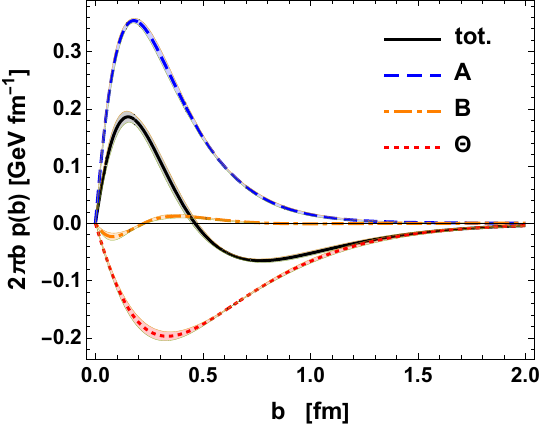}
    \caption{Various contributions to the transverse pressure (multiplied with $2\pi b$) according to Eq.~(\ref{eq:pdec}). The repulsive part comes from $A$, the confining part from $\Theta$, and the contribution of $B$ is small and changes the sign. \label{fig:pdec}}
\end{figure}

Finally, we note that the pressure can be written as 
\begin{eqnarray}
&& p(b)=\frac{m_N}{6}A(b)+ \frac{1}{24 m_N}\frac{1}{b} \frac{d}{db}b \frac{d}{db}B(b) - \frac{1}{6}\Theta(b). \nonumber \\
\label{eq:pdec}
\end{eqnarray}
This decomposition is shown in Fig.~\ref{fig:pdec}. We can see that the contribution of $A$, 
which by the general argument is positive, is repulsive and short-range, 
the contribution of $B$ is small, whereas the contribution of $\Theta$ is attractive and long range. Certainly, the long range 
of the $\Theta$ part reflects the smallness of the $\sigma$ mass, while the short range of $A$ is associated with the larger mass of $f_2$. 
Note that this very simple interpretation depends crucially on the proper spin decomposition of SEM introduced in Sec.~\ref{sec:general}.
Qualitatively similar results were obtained in~\cite{Ji:2025gsq,Fujii:2025aip}, where the $0^{++}$ component is associated with the 
gluon contribution to the trace anomaly.

\subsection{Transverse radii}

The two-dimensional transverse radii are obtained from the densities of Sec.~\ref{sec:td} as 
\begin{eqnarray}
\langle b^2 \rangle_F = \frac{\int_0^\infty 2\pi b \, b^2 F(b)}{\int_0^\infty 2\pi b \,  F(b)}=\frac{4}{F(0)} \left . \frac{dF(t)}{dt} \right |_{t=0}, \label{eq:rbg}
\end{eqnarray}
whereas the mechanical radius is
\begin{eqnarray}
\langle b^2 \rangle_{\rm mech} = \frac{\int_0^\infty 2\pi b \, b^2 [p(b)+\tfrac{1}{2}s(b)]}{\int_0^\infty 2\pi b [p(b)+\tfrac{1}{2}s(b)]} 
= \frac{4 D(0)}{\int_0^\infty d(-t) D(t)} . \nonumber \\
\end{eqnarray}
In our meson dominance model we find 
\begin{eqnarray}
&& \langle b^2 \rangle_A=4 \left ( -c_A+\frac{1}{m_{f_2}^2} +\frac{1}{m_{f'_2}^2}+\frac{1}{m_{f''_2}^2}+\frac{1}{m_{f'''_2}^2} \right ) = \nonumber \\
&& \hspace{1.4cm} [0.34(1)~{\rm fm}]^2, 
\end{eqnarray}
where numerically $c_A$ approximately cancels the contribution ${1}/{m_{f''_2}^2}+{1}/{m_{f'''_2}^2}$. Similarly,
\begin{eqnarray}
\langle b^2 \rangle_\Theta=4 \left (\frac{1}{m_{\sigma}^2} +\frac{1}{m_{f_0}^2} \right ) =[0.60(3)~{\rm fm}]^2,
\end{eqnarray}
where the comparatively large size reflects the smallness of the effective $\sigma$ mass.
Finally, 
\begin{eqnarray}
\langle b^2 \rangle_{\rm mech}=[0.48(3)~{\rm fm}]^2,
\end{eqnarray}
with the hierarchy $\langle b^2 \rangle_A<\langle b^2 \rangle_{\rm mech} < \langle b^2 \rangle_\Theta$.

\section{Radial densities}
\label{sec:3d}

\subsection{Definitions}

For $F=A$, $B$, or $\Theta$, the three-dimensional spatial distributions~\cite{Polyakov:2018zvc} are defined in the Breit frame as
\begin{eqnarray}
&& F(r)=\int \frac{d^3\Delta}{(2\pi)^3} e^{-i {\bf \Delta}\cdot {\bf r}} F(-\Delta^2)= \nonumber \\
&& \hspace{1cm} \frac{1}{2\pi^2}\int d\Delta \, \Delta^2 j_0(\Delta r) F(-\Delta^2), \label{eq:Frr}
\end{eqnarray}
where $\Delta^2=-t$ and $j_0(z)=\sin(z)/z$ is the spherical Bessel function. 
We recall the statement of~\cite{Panteleeva:2021iip} that for the nucleon (for other spin states this does not hold~\cite{Freese:2021mzg})
the three-dimensional and the transverse distributions are connected via the Abel transform. To see this explicitly, we notice that the Abel transform of the kernel in~(\ref{eq:Frr}) is
\begin{eqnarray}
2 \int_b^\infty r \Delta \frac{j_0(\Delta r)}{\sqrt{r^2-b^2}} dr = \pi  J_0(\Delta b),
\end{eqnarray}
which yields~(\ref{eq:Fgen}). Therefore the ``tomographic information'' contained in the transverse densities of the nucleon describes fully the three-dimensional distributions, and vice versa.

The case of $J$ is a bit more involved, as the spatial distribution contains both the monopole and quadrupole terms~\cite{Lorce:2017wkb}, 
\begin{eqnarray}
&&J^i(r)=s^i \int \frac{d^3\Delta}{(2\pi)^3} e^{-i {\bf \Delta}\cdot {\bf r}} \left [ J(t)+\frac{2t}{3} \frac{dJ(t)}{dt} \right ]_{t=-\Delta^2} \!\!\!+  \\
&& \hspace{12mm} s^j \int \frac{d^3\Delta}{(2\pi)^3} e^{-i {\bf \Delta}\cdot {\bf r}} 
 (\Delta^i \Delta^j -\tfrac{1}{3} \delta^{ij} \Delta^2) \left . \frac{dJ(t)}{dt} \right |_{t=-\Delta^2}, \nonumber 
\end{eqnarray}
It can be written as
\begin{eqnarray}
\hspace{-5mm} J^i(r)=s^i J^{\rm mon}(r) + s^j \left ( \frac{r^i r^j}{r^2}-\tfrac13 \delta^{ij}  \right ) J^{\rm quad}(r), 
\end{eqnarray}
where the monopole and quadrupole terms are given by
\begin{eqnarray}
 J^{\rm mon}(r) = -\tfrac23 J^{\rm quad}(r) = \frac{r}{6\pi^2} \int_0^\infty d\Delta\,\Delta^3 j_1(\Delta r) J(-\Delta^2). \nonumber \\
\end{eqnarray}
Note the simple proportionality relation with the factor of $-2/3$ between the monopole and quadrupole parts, which is a generic feature when $J(t)$ in nonsingular, allowing for 
integration by parts.

\subsection{Spectral representation}

Using the dispersion relation~(\ref{eq:dis}) in~(\ref{eq:Frr}), in analogy to~(\ref{eq:FbK}), we get the spectral representation of the three-dimensional distributions, 
\begin{eqnarray}
F(r) = \frac{1}{4\pi^2} \int_{4m_\pi^2}^\infty ds \frac{e^{-r \sqrt{s}}}{r} \, {\rm Im}\, F(s). \label{eq:FrK}
\end{eqnarray}
Notice as a check that the Abel transform of~(\ref{eq:FrK}) yields~(\ref{eq:FbK}).

Expanding $\exp( -r \sqrt{s})$ at low $r$ and using the super-convergence sum rule~(\ref{eq:srn}) for $n=1$ we get 
\begin{eqnarray}
F(r=0) = - \frac{1}{4\pi^2} \int_{4m_\pi^2}^\infty ds \sqrt{s} \, {\rm Im}\, F(s),  \label{eq:FrK0}
\end{eqnarray}
which indicates an even stronger sensitivity to large $s$ than in the
two-dimensional case of~(\ref{eq:FbK}).  With the overall sign
in~(\ref{eq:FrK0}), the positivity of $F(0)$ means that there
are parts of the corresponding spectrum with negative strength.  When
Eq.~(\ref{eq:srn}) holds for $n=2$, then
\begin{eqnarray}
F(r) = F(r=0)- \frac{r^2}{24\pi^2} \int_{4m_\pi^2}^\infty ds \, s^{3/2} \, {\rm Im}\, F(s)+{\cal O}(r^3), \nonumber \\ \label{eq:FrK2}
\end{eqnarray}
and additionally with $n=3$
\begin{eqnarray}
&& F(r) = F(r=0) +  F'(r=0)r^2  \nonumber \\
 && \hspace{7mm}+ \frac{r^4}{480\pi^2} \int_{4m_\pi^2}^\infty ds \, s^{5/2} \, {\rm Im}\, F(s)+{\cal O}(r^5). \nonumber \\ \label{eq:FrK4}
\end{eqnarray}
For the monopole angular momentum form factor 
\begin{eqnarray}
\hspace{-5mm} J^{\rm mon}(r) = \frac{1}{12\pi^2} \int_{4m_\pi^2}^\infty ds \frac{e^{-r \sqrt{s}}(r \sqrt{s}+1)}{r} \, {\rm Im}\, J(s). 
\nonumber \\ \label{eq:JrK}
\end{eqnarray}
With the super-convergence sum rules, the lowest term in the low-$r$ expansion is
\begin{eqnarray}
\hspace{-3mm}J^{\rm mon}(r) = \frac{r^2}{36\pi^2} \int_{4m_\pi^2}^\infty ds  \,s^{3/2}  {\rm Im}\, J(s) +{\cal O}(r^3). \label{eq:JrKd}
\end{eqnarray}

\subsection{Radial pressure and shear forces}

\begin{figure}[t]
    \centering
    \includegraphics[width=0.45\textwidth]{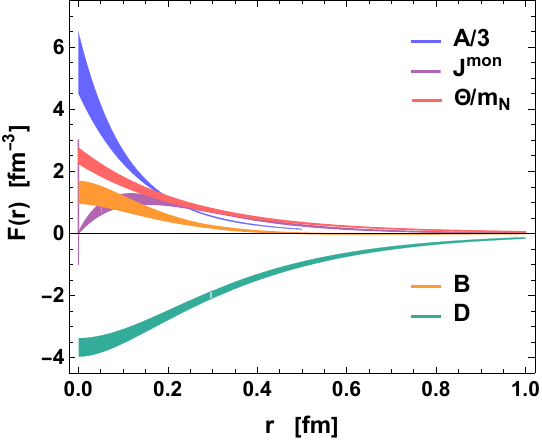}
    \caption{Three-dimensional distributions of $A$ (divided by 3), $J^{\rm mon}$, $\Theta/m_N$, $B$, and $D$, plotted as functions of the radial coordinate $r$, following from the meson dominance model fitted to the MIT lattice data at $m_\pi=170$~MeV. \label{fig:3Dden} }
\end{figure}

From the form of~(\ref{eq:FrK0}-\ref{eq:FrK4},\ref{eq:JrKd}), we expect the three-dimensional distributions near the origin 
 to be sensitive to the details of the high-$s$ spectrum, hence to the particular model used. This is the same issue that occurred in the 
 transverse densities, but now the model dependence is even stronger. 
 
\begin{figure}[t]
    \centering
    \includegraphics[width=0.45\textwidth]{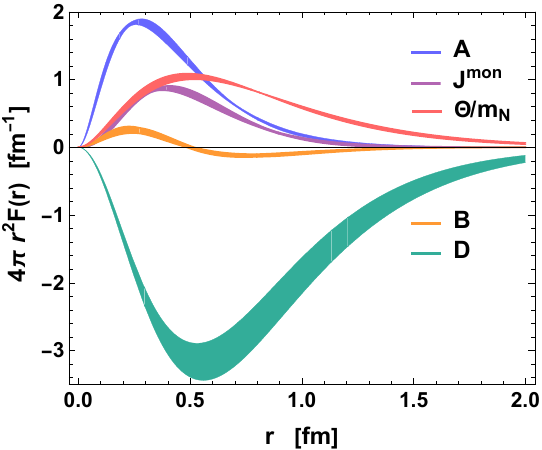}
    \caption{Radial densities of $A$, $J^{\rm mon}$, $\Theta/m_N$, $B$, and $D$, plotted as functions of the radial coordinate $r$, following from the meson dominance model fitted to the MIT lattice data at $m_\pi=170$~MeV. \label{fig:3Ddenr} }
\end{figure}

 In Fig.~\ref{fig:3Dden} we plot the 
 three-dimensional densities for the nucleon from the meson dominance model. We note that at the origin they have a large uncertainty, as discussed above. The density $A(r)$ is the  largest, as the corresponding form factor has a longest tail in $-t$, caused by the large values of the tensor-meson masses. Except for $B(r)$, which changes the sign, all the other densities have a definite sign. The corresponding radial densities are plotted in Fig.~\ref{fig:3Ddenr}.
 
 The three-dimensional pressure and shear are evaluated as~\cite{Polyakov:2018zvc}
 \begin{eqnarray}
&& p(r)=- \frac{1}{12\pi^2 m_N}\int d\Delta \, \Delta^4 j_0(\Delta r) D(-\Delta^2), \nonumber \\ 
&& s(r) =- \frac{1}{8\pi^2 m_N}\int d\Delta \, \Delta^4 j_2(\Delta r) D(-\Delta^2). \label{eq:Fr}
 \end{eqnarray}
 They satisfy all the relations spelled out in~\cite{Polyakov:2018zvc}, in particular $\int_0^\infty 4\pi r^2 p(r) dr=0$ and
 $-m_N \int_0^\infty 4\pi r^4 p(r) dr=-\tfrac{4}{15}m_N  \int_0^\infty 4\pi r^4 s(r) dr = D(0)$. 
 
\begin{figure}[t]
    \centering
    \includegraphics[width=0.45\textwidth]{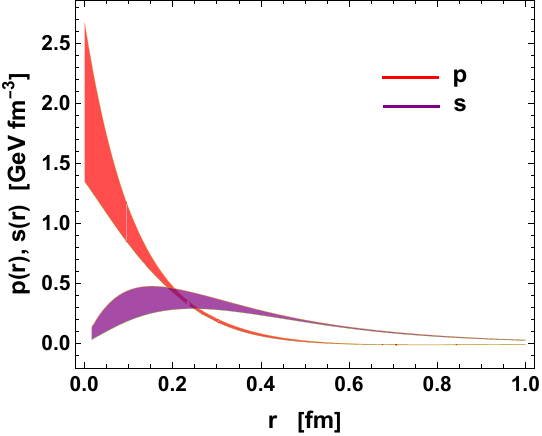}
    \caption{Same as in Fig.~\ref{fig:3Dden} but for the pressure and stress. 
    The pressure changes sign from $+$ to $-$ as $r$ increases, at $r\simeq 0.5$~fm, and $\int_0^\infty 4\pi r^2 p(r) dr=0$, as required by stability. \label{fig:3Dps}}
\end{figure}

\begin{figure}[t]
    \centering
    \includegraphics[width=0.45\textwidth]{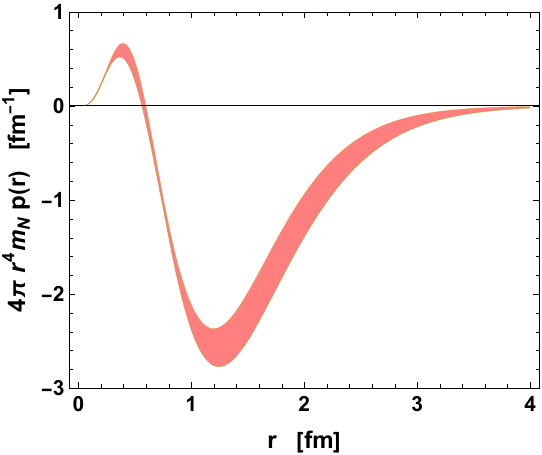}
    \caption{Same as in Fig.~\ref{fig:3Dps} but for the pressure multiplied with $4\pi r^4 m_N$, which upon integration yields $D(0)$.  \label{fig:3Dp4}}
\end{figure}

 We plot these mechanical distributions in Fig.~\ref{fig:3Dps}.
 Higher central pressure than that obtained from the chiral soliton model~\cite{Goeke:2007fp} means that the nucleon is a spatially more compact object in the meson dominance approach. In Fig.~\ref{fig:3Dp4} we plot $4\pi r^4 m_N p(r)$, which integrates to $D(0)$.

Using the same methods as in Sec.~\ref{sec:trp} we get the sum rules
\begin{eqnarray}
&& p(r=0) = -\frac{1}{24\pi^2 m_N} \int_{4m_\pi^2}^\infty ds \, s^{3/2} \, {\rm Im}\, D(s), \\
&& \left . \frac{ds(r)}{dr^2}\right |_{r=0} = \frac{1}{240\pi^2 m_N} \int_{4m_\pi^2}^\infty ds \, s^{5/2} \, {\rm Im}\, D(s).  \nonumber
\end{eqnarray}

The large-$r$ asymptotics at $r\to\infty$, following from the threshold behavior of Eq.~(\ref{eq:delThet}), is 
\begin{eqnarray}
&& p(r) \sim - e^{-2m_\pi r}/r^{5/2}, \nonumber \\
&& s(r) \sim + e^{-2m_\pi r}/r^{5/2}, \label{eq:powr}
\end{eqnarray}
where the tail of $\Theta(r) \sim + e^{-2m_\pi r}/r^{5/2}$ determines the falloff. For completeness, we mention that 
$A(r), B(r) \sim + e^{-2m_\pi r}/r^{9/2}$ and 
$J^{\rm mon}(r) \sim + e^{-2m_\pi r}/r^{7/2}$.
We notice that the power of $1/r$ in the tails of Eq.~(\ref{eq:powr}) differs from the result obtained in the 
Chiral Soliton model~\cite{Goeke:2007fp}, where the falloff was faster, $\sim  e^{-2m_\pi r}/r^{6}$. This is probably caused by 
the large-$N_c$ limit taken first in that approach. Our asymptotics robustly reflects the kinematic behavior at the 
$2\pi$ production threshold.

\subsection{Radii and the $D$-term \label{sec:radii}}

The three-dimensional mean squared radii are
\begin{eqnarray}
\langle r^2 \rangle_F=\frac{\int d^3r \,r^2F(r)}{\int d^3r \,F(r)}=\frac{6}{F(0)} \left . \frac{dF(t)}{dt} \right |_{t=0}. 
\end{eqnarray}
In our meson dominance model
\begin{eqnarray}
&& \langle r^2 \rangle_A=6 \left ( -c_A+\frac{1}{m_{f_2}^2} +\frac{1}{m_{f'_2}^2}+\frac{1}{m_{f''_2}^2}+\frac{1}{m_{f'''_2}^2} \right ) = \nonumber \\
&& \hspace{1.4cm} [0.51(1)~{\rm fm}]^2, 
\end{eqnarray}
and
\begin{eqnarray}
&& \langle r^2 \rangle_\Theta=6 \left (\frac{1}{m_{\sigma}^2} +\frac{1}{m_{f_0}^2} \right ) =[0.90(4)~{\rm fm}]^2.
\end{eqnarray}
For the angular momentum density
\begin{eqnarray}
&& \langle r^2 \rangle_J=\frac{\int d^3r \,r^2J^i(r)}{\int d^3r \,J^i(r)}=\frac{10}{J(0)} \left . \frac{dJ(t)}{dt} \right |_{t=0} = \nonumber \\
&& \hspace{12mm} 20\left . \frac{dJ(t)}{dt} \right |_{t=0}, 
\end{eqnarray}
with only the monopole term contributing. In the meson dominance model 
\begin{eqnarray}
&& \langle r^2 \rangle_J=10 \left ( -c_J+\frac{1}{m_{f_2}^2} +\frac{1}{m_{f'_2}^2}+\frac{1}{m_{f''_2}^2}+\frac{1}{m_{f'''_2}^2} \right ) = \nonumber \\
&& \hspace{1.4cm} [0.57(3)~{\rm fm}]^2. 
\end{eqnarray}
The  $D$-term is related to the mean square radii as follows:
\begin{eqnarray}
D(0)=\frac{2m_N^2}{9} \left (  \langle r^2 \rangle_A - \langle r^2 \rangle_\Theta   \right ).
\end{eqnarray}
The mechanical radius, as defined in~\cite{Polyakov:2018zvc}, is
\begin{eqnarray}
\langle r^2 \rangle_{\rm mech} = \frac{6 D(0)}{\int_0^\infty d(-t) D(t)} = [0.72(5)~{\rm fm}]^2, 
\end{eqnarray}
and the energy radius, defined via $T^{00}$~\cite{Polyakov:2018zvc}, is 
\begin{eqnarray}
&& \langle r^2 \rangle_E= \langle r^2 \rangle_A -\frac{3}{2m_N^2} D(0) =  \\
&& \hspace{1cm}\frac{2}{3}  \langle r^2 \rangle_A +\frac{1}{3}  \langle r^2 \rangle_\Theta = [0.67(2)~{\rm fm}]^2 \nonumber 
\end{eqnarray}
(cf. various estimates for $\langle r^2 \rangle_{\rm mass} \equiv \langle r^2 \rangle_E$ collected in~\cite{Goharipour:2025yxm}).
To summarize, we find the hierarchy
\begin{eqnarray}
\langle r^2 \rangle_A < \langle r^2 \rangle_J < \langle r^2 \rangle_E < \langle r^2 \rangle_{\rm mech}  < \langle r^2 \rangle_\Theta,
\end{eqnarray}
spanning the sizes from 0.5~fm to 0.9~fm.
Note that the largest value of $\langle r^2 \rangle_\Theta$ reflects the fact that the trace-anomaly lattice 
form factor of Fig.~\ref{fig:Th} is the longest range, and hence has the highest slope at the origin, essentially reflecting 
the smallness of the effective $\sigma$ mass. This is in 
contrast to the extraction by Kharzeev~\cite{Kharzeev:2021qkd} from the GlueX $J$/$\psi$ photoproduction data~\cite{GlueX:2019mkq}, where a 
relatively small value of $\langle r^2 \rangle_\Theta\sim [0.55~{\rm fm}]^2$ was found (see also the discussion in~\cite{Du:2020bqj,JointPhysicsAnalysisCenter:2023qgg}).

\section{Conclusions}

In this paper we have discussed several general issues concerning GFFs
of the nucleon and its mechanical properties, involving dispersion
relations, super-convergence sum rules, asymptotic behavior, or
positivity, as well as offered a natural fit of the available lattice
data within the meson dominance approach.

On the general grounds, GFFs of the nucleon can be scrutinized from
the point of view of analyticity realized through the established long
ago meson dominance principle, and implemented through the pertinent
current-field identities. While GFFs are admittedly complicated
structures for the time-like momenta, displaying resonance profiles
corresponding to the exchanged mesons and the corresponding
backgrounds, the space-like region, which has been lately the main
concern both for the theory and experiment, is largely independent of
these details.  We have shown that the meson dominance works well for
the lattice QCD data for the nucleon in the available space-like
domain, mirroring the success for the pion~\cite{Broniowski:2024oyk}.

Since the pQCD constraints impose known power falloffs based on the
leading two-gluon exchange process, $A, J \sim \alpha^2/(-t)^2$, $B, D \sim
\alpha^2/(-t)^3$,  these requirements have been incorporated into our model in terms of
a minimum number of resonances. The $Q^2$ range covered by the lattice QCD simulation, $0
\le Q^2 \le 2~{\rm GeV}^2$, does not probe precisely the asymptotic
fall-off, therefore the corresponding super-convergence sum rules, or the
spectral representation of the spatial densities, are only weakly constrained by the present data. 

Estimates for various radii, exhibiting a certain hierarchy,  and for the value of the $D$-term, have been made within the meson dominance model. 
A large value of the mass (trace-anomaly) radius, reflecting the
smallness of the effective $\sigma$ mass, has been found from the fit. 
A very similar best-fit value, $m_\sigma \sim 0.65~{\rm GeV}$, has been found for both the pion and the nucleon.

For the spatial distributions, we have derived sum rules for their values and derivatives at the origin in terms of the spectral integrals with 
the application of the super-convergence sum rules. We argue that these quantities are weakly constrained, since we do not have sufficient knowledge 
on the corresponding spectral strengths. 

From the threshold behavior in
the time-like region for $\pi\pi \to N\bar{N}$,  using Watson's theorem, we have 
derived the leading asymptotic behavior at large values of the coordinate for the spatial densities, including the pressure and shear forces. 
These formulas involve the obvious two-pion Yukawa tail, multiplied with a power falloff. 
The formulas obtained from our study 
exhibit the signs following from the $\pi-N$ elastic scattering data, which are positive for all the considered densities, except for the pressure, where the coefficient is negative.

The decomposition of SEM into the $0^{++}$ and $2^{++}$ parts offers a natural explanation of the spatial variance of pressure, whose 
attractive tail is associated with the small mass of the $\sigma$ meson, while the inside repulsion is due to the large mass of the $f_2$ meson. 

We also bring up the result~\cite{Broniowski:2024mpw} that the transverse distribution related to the $T^{++}$ is positive at an operator level. 
This implies in general that $A(b)>0$, which is satisfied in our model. 

\bigskip

We are grateful to the authors of Refs.~\cite{Hackett:2023rif}
and~\cite{Wang:2024lrm} for sending us the numbers for their lattice
results. One of us (ERA) thanks Christian Weiss for clarifications
regarding the positivity conditions.

ERA was supported by
Spanish MINECO and European FEDER funds grant and by Project No.
PID2023-147072NB-I00 funded by MCIN/AEI/10.13039- 501100011033,
and by Junta de Andaluc\'\i a Grant No. FQM-225.

\appendix

\section{Features of the minimum meson dominance ansatz \label{app:fit}}

Expansion of the meson-dominance function for $A(t)$ from  Eq.~(\ref{eq:AJmod}) at low $t$ yields
\begin{eqnarray}
A(t)=1+St+Rt^2+{\cal O}(t^3),
\end{eqnarray}
with
\begin{eqnarray}
&& S=-c_A+\frac{1}{m_{f_2}^2}+\frac{1}{m_{f'_2}^2}+\frac{1}{m_{f''_2}^2}+\frac{1}{m_{f'''_2}^2}, \\
&& R= c_2-\frac{c_A^2}{2}+\frac{1}{2}\left ( \frac{1}{m_{f_2}^4}+\frac{1}{m_{f'_2}^4}+\frac{1}{m_{f''_2}^4}+\frac{1}{m_{f'''_2}^4} +S^2 \right).  \nonumber
\end{eqnarray}
Thus, for a given set  of the four resonance masses, $c_A$ and $c_2$ can be adjusted to fit the slope and the curvature of the function at the origin.
At asymptotically large $-t$,
\begin{eqnarray}
A(t)=\frac{c_2 m_{f_2}^2 m_{f'_2}^2 m_{f''_2}^2 m_{f'''_2}^2}{(-t)^2}+{\cal O}[1/(-t)^3].
\end{eqnarray}
Analogous formulas hold for $J(t)$.

For $B(t)$, we have the features
\begin{eqnarray}
&& B(t)=(c_A\!-\!c_J) \left [ t +\left (\frac{1}{m_{f_2}^2}+\frac{1}{m_{f'_2}^2}+\frac{1}{m_{f''_2}^2}+\frac{1}{m_{f'''_2}^2} \right )t^2 \right] \nonumber \\
&& \hspace{1cm} + {\cal O}(t^3),
\end{eqnarray}
and
\begin{eqnarray}
\hspace{-7mm} B(t)=\frac{(c_J-c_A) m_{f_2}^2 m_{f'_2}^2 m_{f''_2}^2 m_{f'''_2}^2}{(-t)^3}+{\cal O}[1/(-t)^4].
\end{eqnarray}

\bibliography{Ref,refs-barpi,newrefs}

\end{document}